\documentclass[10pt,journal,compsoc]{IEEEtran}
\usepackage{ colortbl}
\usepackage{ xcolor}
\usepackage{cite}
\usepackage{multirow}
\usepackage{subfigure}
\usepackage{amssymb}
\usepackage{xspace}
\usepackage{epsf}
\usepackage{setspace}
\usepackage{caption}
\usepackage{comment}
\usepackage{color}
\usepackage{booktabs}
\usepackage{algpseudocode} 
\usepackage{balance}
\usepackage{url,epsfig,array}
\usepackage{amsmath}
\usepackage{epstopdf}
\usepackage{cases}
\usepackage{bm}
\usepackage{color}
\usepackage{tikz}
\usepackage{threeparttable}
\usepackage{pifont}
\usepackage[linesnumbered,ruled,vlined]{algorithm2e}  

\newcommand*{\circled}[1]{\lower.7ex\hbox{\tikz\draw (0pt, 0pt)%
    circle (.5em) node {\makebox[1em][c]{\small #1}};}}

\def\ie{\textit{i.e.}\xspace}
\def\etal{\textit{et al.}\xspace}

\def\eg{\textit{e.g.}\xspace}

\begin{document}
\title{\huge{FedQuad: Adaptive Layer-wise LoRA Deployment and  Activation Quantization for Federated Fine-Tuning}}
 \author{
         Rukuo Li, Jianchun Liu, Hongli Xu, Liusheng Huang
 	\IEEEcompsocitemizethanks{
 		\IEEEcompsocthanksitem R. Li, J. Liu, H. Xu, L. Huang are with the School of Computer Science and Technology, University of Science and Technology of China, Hefei, Anhui, China, 230027, and also with Suzhou Institute for Advanced Research, University of Science and Technology of China, Suzhou, Jiangsu, China, 215123. E-mails: xxx 
 	}
 }

\markboth{IEEE Transactions on xxx}%
{Shell \MakeLowercase{\textit{et al.}}: Bare Advanced Demo of IEEEtran.cls for Journals}

\IEEEtitleabstractindextext{
\begin{abstract}

Federated fine-tuning (FedFT) provides an effective paradigm for fine-tuning large language models (LLMs) in privacy-sensitive scenarios. 
However, practical deployment remains challenging due to the limited resources on end devices. 
Existing methods typically utilize parameter-efficient fine-tuning (PEFT) techniques, such as Low-Rank Adaptation (LoRA), to substantially reduce communication overhead. 
Nevertheless, significant memory usage for activation storage and computational demands from full backpropagation remain major barriers to efficient deployment on resource-constrained end devices. Moreover, substantial resource heterogeneity across devices results in severe synchronization bottlenecks, diminishing the overall fine-tuning efficiency.
To address these issues, we propose FedQuad, a novel LoRA-based FedFT framework that adaptively adjusts the LoRA depth (the number of consecutive tunable LoRA layers from the output) according to device computational capabilities, while employing activation quantization to reduce memory overhead, thereby enabling efficient deployment on resource-constrained devices. Specifically, FedQuad first identifies the feasible and efficient combinations of LoRA depth and the number of activation quantization layers based on device-specific resource constraints. 
Subsequently, FedQuad employs a greedy strategy to select the optimal configurations for each device, effectively accommodating system heterogeneity. 
Extensive experiments demonstrate that FedQuad achieves a 1.4–5.3$\times$ convergence acceleration compared to state-of-the-art baselines when reaching target accuracy, highlighting its efficiency and deployability in resource-constrained and heterogeneous end-device environments.
\end{abstract}
}

\maketitle
\IEEEdisplaynontitleabstractindextext
\IEEEpeerreviewmaketitle

\section{Introduction}\label{sec:intro}
Recent advances in large language models (LLMs), such as GPT-4 \cite{achiam2023gpt} and DeepSeek-V3 \cite{liu2024deepseek}, have demonstrated remarkable modeling capabilities across various natural language processing (NLP) tasks. 
Serving as foundational models, LLMs naturally stimulate interest in fine-tuning for downstream domain-specific tasks, including question answering \cite{jiang2020can}, sentiment analysis \cite{zhang2023sentiment}, and machine translation \cite{zhang2023prompting}. 
However, centralized fine-tuning methods rely on aggregating substantial amounts of domain-specific data from end devices, posing privacy risks \cite{huang2024keystrokesniffer} and facing constraints imposed by regulations such as GDPR \cite{voigt2017eu}.
To address these concerns, federated fine-tuning (FedFT), a distributed fine-tuning approach, has emerged \cite{lin2021fednlp, zhang2023fedpetuning}. FedFT enables end devices to perform local model fine-tuning without sharing raw data, subsequently aggregating model parameters from individual devices at a parameter server (PS) to facilitate knowledge sharing across devices. 

Nevertheless, the practical deployment of FedFT faces substantial challenges due to the mismatch between the enormous scale of LLMs and the limited resources (\ie, computational power, communication bandwidth, and memory) of end devices. For instance, standard NLP LLMs such as RoBERTa-large \cite{radford2019language} demand approximately 4 GB of network traffic per training round and over 34,00 TFLOPs of computation. 
In contrast, typical end devices, such as the NVIDIA Jetson TX2 \cite{mittal2019survey}, provide less than 2 TFLOPS of computational power.  In addition, the available bandwidth for end devices is typically below 100 Mbps,  which is characteristic of typical WAN environments  \cite{liu2023finch}. 
As a result, the convergence process in real-world scenarios may extend to hundreds of hours \cite{cai2022fedadapter}. Furthermore, fine-tuning RoBERTa-large requires around 25 GB of memory, exceeding the capabilities of most end devices, which typically have less than 16 GB of memory \cite{androidauthority2025}.
Recent studies on FedFT have primarily mitigated communication overhead by leveraging various parameter-efficient fine-tuning (PEFT) methods \cite{cai2022fedadapter, zhang2023fedpetuning}, such as Adapter \cite{houlsby2019parameter} and Low-Rank Adaptation (LoRA) \cite{hu2021lora}. Specifically, LoRA reduces the number of trainable parameters by freezing the base model and updating only trainable low-rank matrices inserted into transformer layers~\cite{hu2021lora}. As these LoRA modules typically comprise less than 1\% of the total parameters of the base LLM, LoRA substantially reduces communication overhead in FedFT. 

Despite the substantial reduction in communication overhead enabled by LoRA, FedFT still faces two major challenges, \ie, resource constraints and system heterogeneity.
\textit{First}, resource (\ie, memory, computing power) constraints remain a critical bottleneck for deploying LLMs on end devices.
We notice that LoRA does not fundamentally eliminate memory overhead, as fine-tuning LLMs still demands substantial memory to store intermediate activations. 
For example, fine-tuning RoBERTa-large \cite{liu2019roberta} with LoRA still demands about 17.4 GB of memory, of which 11.5 GB is devoted to activation storage. Modern end devices typically have only 4–16 GB of RAM \cite{androidauthority2025}, exacerbating memory constraints. 
Additionally, due to the heavy computational demand of fine-tuning, resource-limited devices often exhibit extremely slow training speeds.


\textit{Second}, participating devices typically possess heterogeneous system characteristics. In particular, they show significant disparities in computational capabilities (\eg, CPU frequency), with performance gaps often exceeding tenfold, causing synchronization delays as strong devices wait for weak ones during each training round \cite{liu2025adaptive}. For instance, the FLOPS capability of an iPhone 16 is only about 12\% of that of a desktop GPU RTX 3090. 
Such imbalances often cause devices with weak capabilities to become stragglers, prolonging overall training time and impairing fine-tuning efficiency. 
Furthermore, compared to strong devices, weaker devices may be unable to load the full model and are thus either constrained to smaller sub-models or excluded from the training process altogether, leading to diminished overall performance. 

Existing methods for addressing these challenges are typically classified into two major categories.
The \textit{first} category \cite{su2024fedra}\cite{liu2022no} fine-tunes LoRA parameters on a subset of transformer layers and discards the rest. 
For example, Su \etal \cite{su2024fedra} propose FedRA, which constructs sub-models by randomly selecting transformer layers to accommodate resource constraints. 
Liu \etal \cite{liu2022no} introduce InclusiveFL, assigning consecutive layers from the input based on device capabilities and employing momentum distillation to improve shallow model performance. 
However, these methods directly discard certain transformer layers, compromising the original model architecture, thereby limiting feature extraction capabilities and degrading accuracy and convergence speed \cite{woo2024dropbp}. 
Besides, memory constraints may prevent stronger devices from fully utilizing their computational resources, forcing them to wait for weaker ones and causing considerable synchronization delays, which degrade overall fine-tuning efficiency.
The \textit{second} category \cite{sun2024exploring}\cite{cho2023heterogeneous} retains all transformer layers while fine-tuning only a subset of LoRA parameters within them, such as by freezing some LoRA layers or reducing the LoRA rank.
For instance, Sun \etal \cite{sun2024exploring} propose selectively fine-tuning critical layers while freezing others to save resources, thereby avoiding accuracy degradation from discarded layers. 
However, the uneven distribution of resource consumption across transformer layers limits the practical deployability of this approach and leads to longer waiting times. 
To enhance fine-tuning efficiency, Cho \etal \cite{cho2023heterogeneous} introduce HetLoRA, assigning different LoRA ranks based on device capabilities to alleviate system heterogeneity. 
Nonetheless, merely adjusting LoRA rank does not fundamentally resolve LoRA’s high computational overhead,
and memory savings remain limited, hindering practical deployment.

In this work, we propose FedQuad, a novel federated fine-tuning framework that adaptively determines the configuration of LoRA depth (\ie, the number of consecutive unfrozen LoRA layers from the output) and the number of activation quantization layers to address challenges arising from resource constraints and system heterogeneity.
As discussed in Section~\ref{sec:motivation}, increasing LoRA depth generally improves fine-tuning performance but also increases resource consumption, making deployment on memory-limited devices impractical. 
Notably, freezing layers beyond the first fine-tuned LoRA layer does little to reduce memory usage. While activation checkpointing reduces memory cost, it adds substantial computational overhead.
To this end, FedQuad proposes to reduce memory usage via activation quantization, thereby enabling deeper LoRA configurations and improving fine-tuning performance. 
Specifically, FedQuad first efficiently identifies all feasible configurations of LoRA depth and quantization layers that satisfy device-specific memory constraints. FedQuad then selects the optimal configuration by jointly considering device computational capability, thereby striking a balance between training efficiency and model accuracy. 
Crucially, FedQuad maintains the full transformer architecture, ensuring both high model fidelity and practical deployability in heterogeneous federated environments.

Even within memory-feasible configurations, the interplay between LoRA depth and activation quantization has a significant impact on overall fine-tuning efficiency.
Increasing the number of quantized layers allows for deeper LoRA integration, thereby enhancing model accuracy.
It also introduces additional computational overhead, potentially leading to prolonged synchronization delays across devices. 
Conversely, reducing the number of quantized layers lowers computational latency but restricts the allowable LoRA depth, ultimately limiting the model's representational capacity and fine-tuning performance.
Therefore, under resource constraints, identifying an optimal trade-off between LoRA depth and the number of quantized layers remains a non-trivial yet essential task for balancing fine-tuning accuracy and efficiency. 

Our main contributions are summarized as follows:
\begin{itemize}
\item We propose FedQuad, an efficient LoRA-based FedFT framework that effectively addresses resource constraints and system heterogeneity by jointly determining the LoRA depth and the number of activation quantization layers.
\item We conduct a comprehensive analysis of the joint impact of LoRA depth and activation quantization on fine-tuning performance.
Based on this insight, we design a greedy-based algorithm that dynamically selects optimal configurations according to device capabilities. 
\item Extensive experimental evaluations demonstrate that our proposed algorithm achieves $1.4$–$5.3\times$ acceleration when achieving the target accuracy, compared to existing solutions.
\end{itemize}

\section{Background and Motivation}\label{sec:motivation}

\subsection{Federated Fine-Tuning LLMs with LoRA}
\textbf{LoRA for LLMs.} With the rapid increase in the number of parameters in LLMs, traditional training methods, such as fine-tuning all model parameters, achieve excellent performance but are extremely resource-intensive and costly.
Consequently, PEFT methods have emerged as essential approaches for the efficient deployment of LLMs. Among various PEFT methods, LoRA has gained widespread popularity as an effective and efficient strategy. LoRA reduces the number of trainable parameters by introducing low-rank adapters, enabling scalable and cost-effective fine-tuning of LLMs.
Specifically,  instead of updating the full parameter matrix of the pre-trained model, LoRA inserts two low-rank matrices, denoted as $\mathcal{A}$ and $\mathcal{B}$, to approximate the weight update. Let $\mathcal{W}_0$ denote the frozen weight matrix of the pre-trained model. The LoRA update can be formulated as: \vspace{-0.1cm} \begin{equation} \mathcal{W}_0 + \Delta \mathcal{W} = \mathcal{W}_0 + \mathcal{B} \mathcal{A} \vspace{-0.1cm} \end{equation} where $\mathcal{W}_0 \in \mathbb{R}^{d \times k}$, $\mathcal{B} \in \mathbb{R}^{d \times r}$ and $\mathcal{A} \in \mathbb{R}^{r \times k}$, with $r \ll \min(d, k)$. 
During training, inputs are simultaneously passed through both the frozen pre-trained model and the LoRA adapters. The outputs are then combined to form the final representation. For an input feature $x_i$ in the $i$-th layer, the output $h_i$ is computed as: \vspace{-0.1cm} \begin{equation} h_i = \mathcal{W}_i x_i + \Delta \mathcal{W}_i x_i = \mathcal{W}_i x_i + \mathcal{B}_i \mathcal{A}_i x_i \vspace{-0.1cm} \end{equation} Only the parameters $\mathcal{A}_i$ and $\mathcal{B}_i$ are updated via gradient descent by minimizing the loss between the model output and the ground-truth labels. Therefore, LoRA significantly reduces  the number of trainable parameters (typically less than 1\% \cite{zhang2023fedpetuning}) while effectively preserving the generalization capability of the pre-trained model.

\textbf{Federated Learning with LoRA.} Federated learning (FL) is a distributed training paradigm that enables multiple end devices to collaboratively train a shared global model without sharing their local data. The global training objective is to find the optimal model parameters 
$\bm{\omega}^* = \{ \Delta \bm{\omega}^*, \bm{\omega}_0 \}$ that minimize the overall loss: 
\vspace{-0.2cm} 

\begin{equation}
\bm{\omega}^* = \operatorname*{argmin}_{\bm{\omega} = \{ \Delta \bm{\omega}, \bm{\omega}_0 \}} f(\bm{\omega})
= \frac{1}{n} \sum_{i=1}^{n} \frac{1}{|\mathbb{D}_i|} \sum_{\xi_i \in \mathbb{D}_i} F_i(\bm{\omega}; \xi_i)
\end{equation}
where $F_i(\bm{\omega}_i; \xi_i)$ is the local loss function computed on device $i$ with data sample $\xi_i$ from its local dataset $\mathbb{D}_i$, and $n$ is the total number of participating end devices.

In each local training round $h$, end device $i$ updates its local parameters using stochastic gradient descent (SGD)~\cite{robbins1951stochastic}: \vspace{-0.1cm} \begin{equation} \Delta \hat{\bm{\omega}}_i^{h} = \Delta \bm{\omega}_i^{h} - \eta \cdot \nabla f_i(\Delta \bm{\omega}_i^{h}) \vspace{-0.1cm} \end{equation} where $\eta$ is the learning rate, and $\nabla f_i(\cdot)$ denotes the gradient of the local loss function with respect to the trainable parameters. After local training, each end device sends the updated gradients $\Delta \bm{\omega}_i^h$ to the central parameter server (PS), which performs model aggregation: \vspace{-0.1cm} \begin{equation} \Delta \bm{\omega}^{h+1} = \frac{1}{n} \sum_{i=1}^{n} \Delta \bm{\omega}_i^h \vspace{-0.1cm} \end{equation} The server then distributes the aggregated updates to all end devices for the next training round.

In conventional FL, the entire model's parameters must be synchronized between end devices and the server, which incurs high communication overhead, particularly for LLMs. When incorporating LoRA into FL, the pre-trained backbone $\bm{\omega}_0$ remains frozen throughout training, and only the low-rank adapter parameters $\Delta \bm{\omega}$ are updated and exchanged. This design substantially reduces the volume of transmitted data on resource-constrained devices, thus accelerating convergence and improving the feasibility of federated training with LLMs.
\subsection{Resource Bottlenecks of Fine-tuning with LoRA }
\label{lora-limitations}
\begin{figure}[t]
    \centering
    \includegraphics[width=0.9\linewidth]{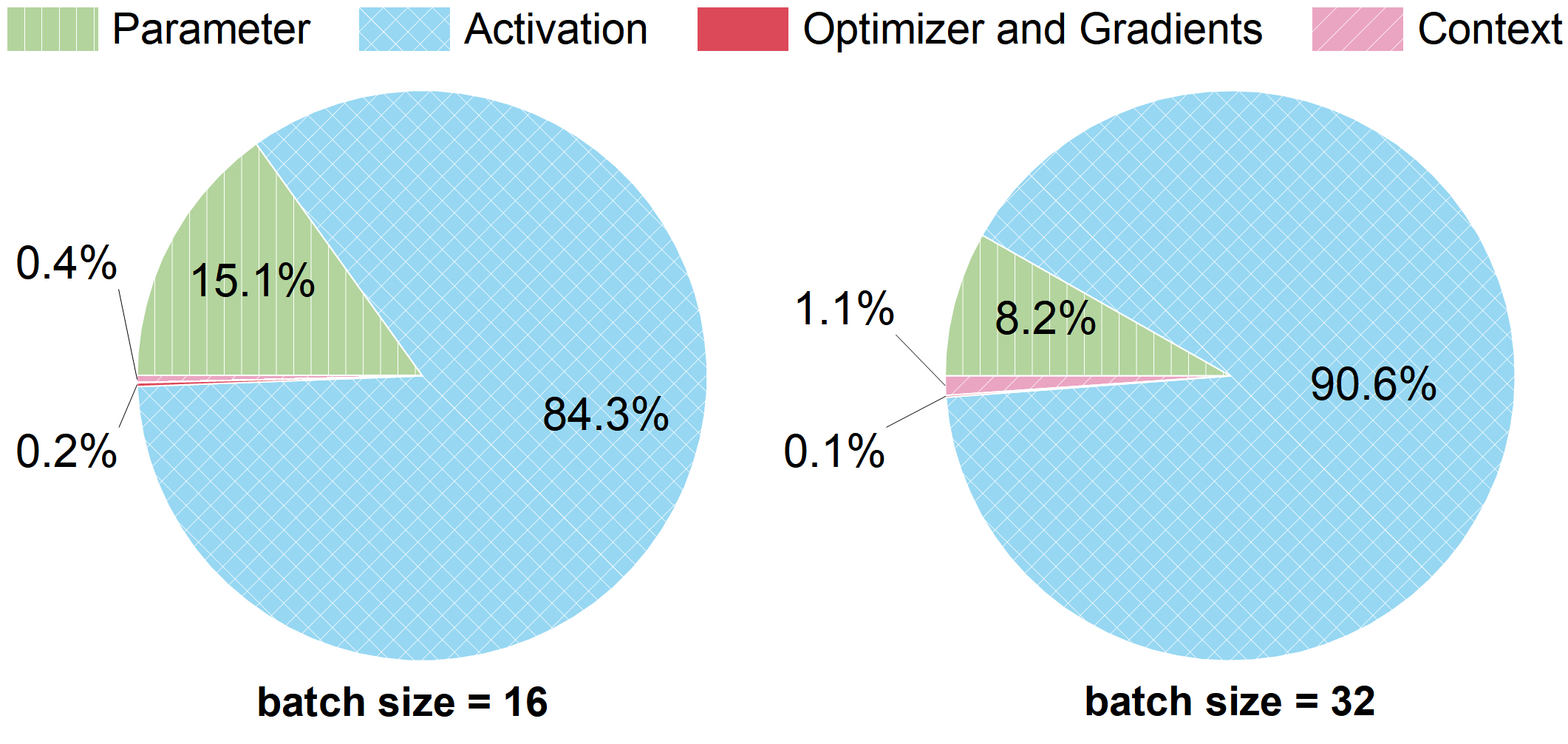}
    \caption{GPU Memory Footprint for LoRA Fine-Tuning.}
    \label{ob-1}
    \vspace{-0.7cm}
\end{figure}
One of the key advantages of LoRA is its ability to significantly reduce the number of parameters that need to be updated during training. By introducing trainable low-rank matrices on top of the original weight matrices, LoRA constrains parameter updates to low-rank matrices, thereby greatly reducing  communication overhead. For instance, when fine-tuning RoBERTa-large, LoRA reduces the per-round communication cost per device from 1355 MB to just 6 MB, accounting for less than 0.5\% of that in full fine-tuning.

Nevertheless, LoRA does not fully address the memory and computation bottleneck during training. First, LoRA still demands substantial memory resources. Although reducing the number of trainable parameters alleviates memory usage for optimizer states and gradient storage, activations still need to be retained for gradient computation during the backward pass, which occupies a substantial portion of GPU memory. In particular, existing LoRA-based federated fine-tuning frameworks, such as FedLoRA~\cite{wu2023fedlora} and HetLoRA~\cite{cho2023heterogeneous}, typically update LoRA modules across all transformer layers. Consequently, activations from every layer must be preserved throughout training, exacerbating memory pressure and limiting the scalability of federated fine-tuning on memory-constrained devices.
To further illustrate the severity of memory consumption caused by activations, we conduct a series of fine-tuning experiments on RoBERTa-large~\cite{liu2019roberta} using the MNLI dataset~\cite{wang2018glue}. The batch size is set to 16 and 32, respectively, with a maximum sequence length of 256, and GPU memory usage is continuously monitored throughout training. As shown in Fig.\ref{ob-1}, activations account for the majority of memory consumption in LoRA-based fine-tuning. With a batch size of 16, activations occupy 85.1\% of total memory usage. When the batch size increases to 32, the total memory consumption reaches 17.4GB, with activation memory usage rising to 15.9~GB, accounting for 91.5\% of the total.
Therefore, activation storage is the primary contributor to the memory bottleneck in LoRA-based fine-tuning.
Therefore, activation storage is the main cause of the memory bottleneck in LoRA-based fine-tuning.

Furthermore, computational power remains a critical bottleneck for LoRA. For instance, fine-tuning RoBERTa-large using LoRA on an A6000 GPU reduces per-batch latency from 1002 ms to approximately 699 ms, which amounts to only a 30.2 \% reduction in computational time compared to full fine-tuning.
Although LoRA dramatically reduces the number of trainable parameters and thus shortens gradient computation and optimizer update time, the forward and backward passes through the frozen backbone still remain computationally intensive. 
 Consequently, resource-constrained devices within FedFT systems experience slow training speeds, resulting in significant synchronization delays and reduced overall fine-tuning efficiency. In summary, while LoRA substantially reduces communication costs, both memory and computational limitations continue to pose significant challenges for LoRA-based fine-tuning of LLMs on end devices with limited resources. 
 


\subsection{Effect of LoRA Depth on Overall Performance }
\label{lora-depth}

\begin{figure}[t]
    \centering
    \subfigure[ Resource overhead]{
        \begin{minipage}[t]{0.5\linewidth}
        \centering
        \includegraphics[width=1.7in]{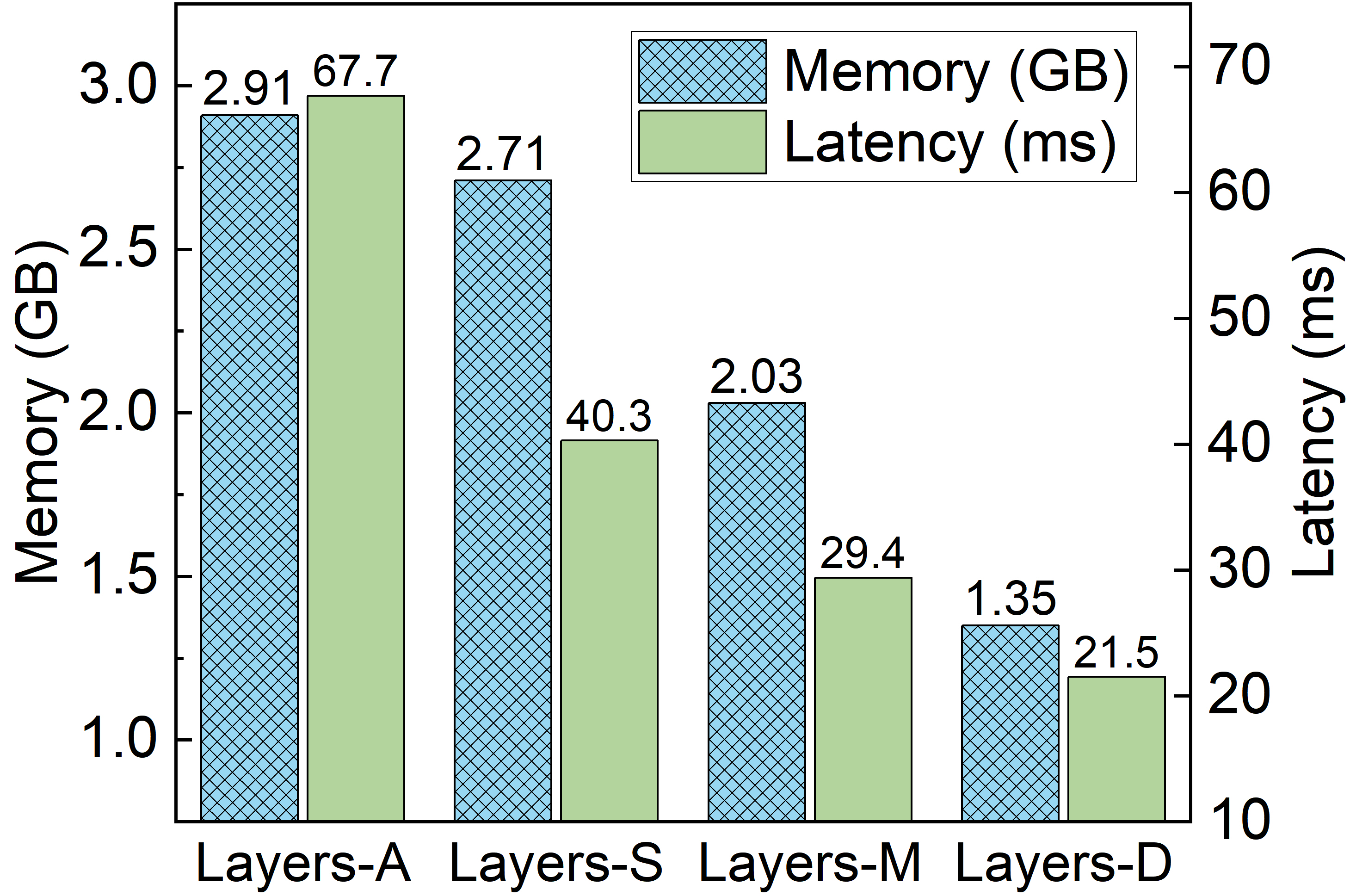}
        \end{minipage}%
        \label{2a}
    }%
    \subfigure[Accuracy]{
        \begin{minipage}[t]{0.5\linewidth}
        \centering
        \includegraphics[width=1.5in]{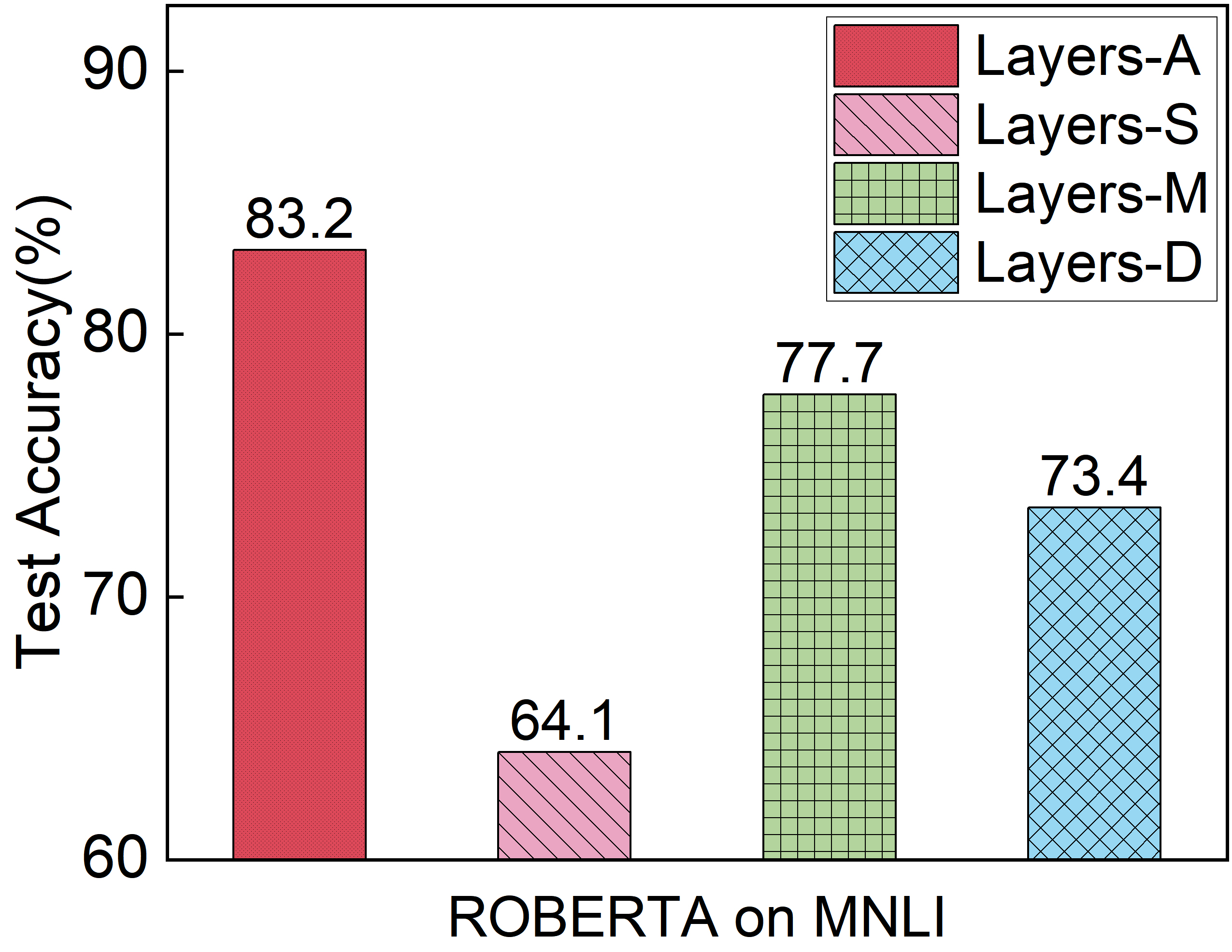}
        \end{minipage}%
        \label{2b}
    }%
    \centering
    \vspace{-0.5cm}
    \caption{ The impact of LoRA position on fine-tuning.}
    \label{ob-2}
    \vspace{-0.6cm}
\end{figure}
\begin{figure}[t]
    \centering
    \subfigure[ Accuracy \& Memory usage]{
        \begin{minipage}[t]{0.5\linewidth}
        \centering
        \includegraphics[width=1.7in]{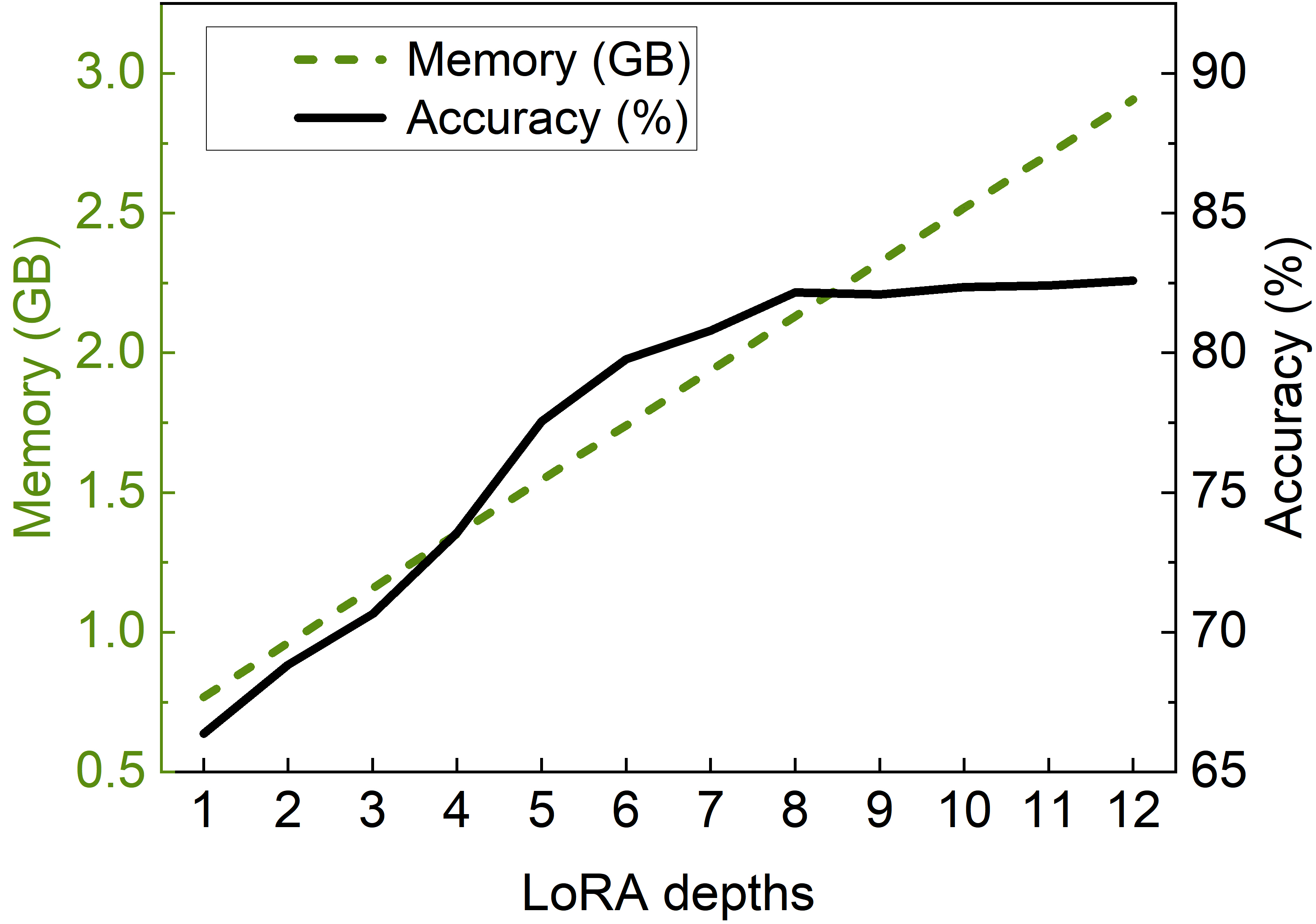}
        \end{minipage}%
        \label{3a}
    }%
    \subfigure[Computational latency]{
        \begin{minipage}[t]{0.5\linewidth}
        \centering
        \includegraphics[width=1.6in]{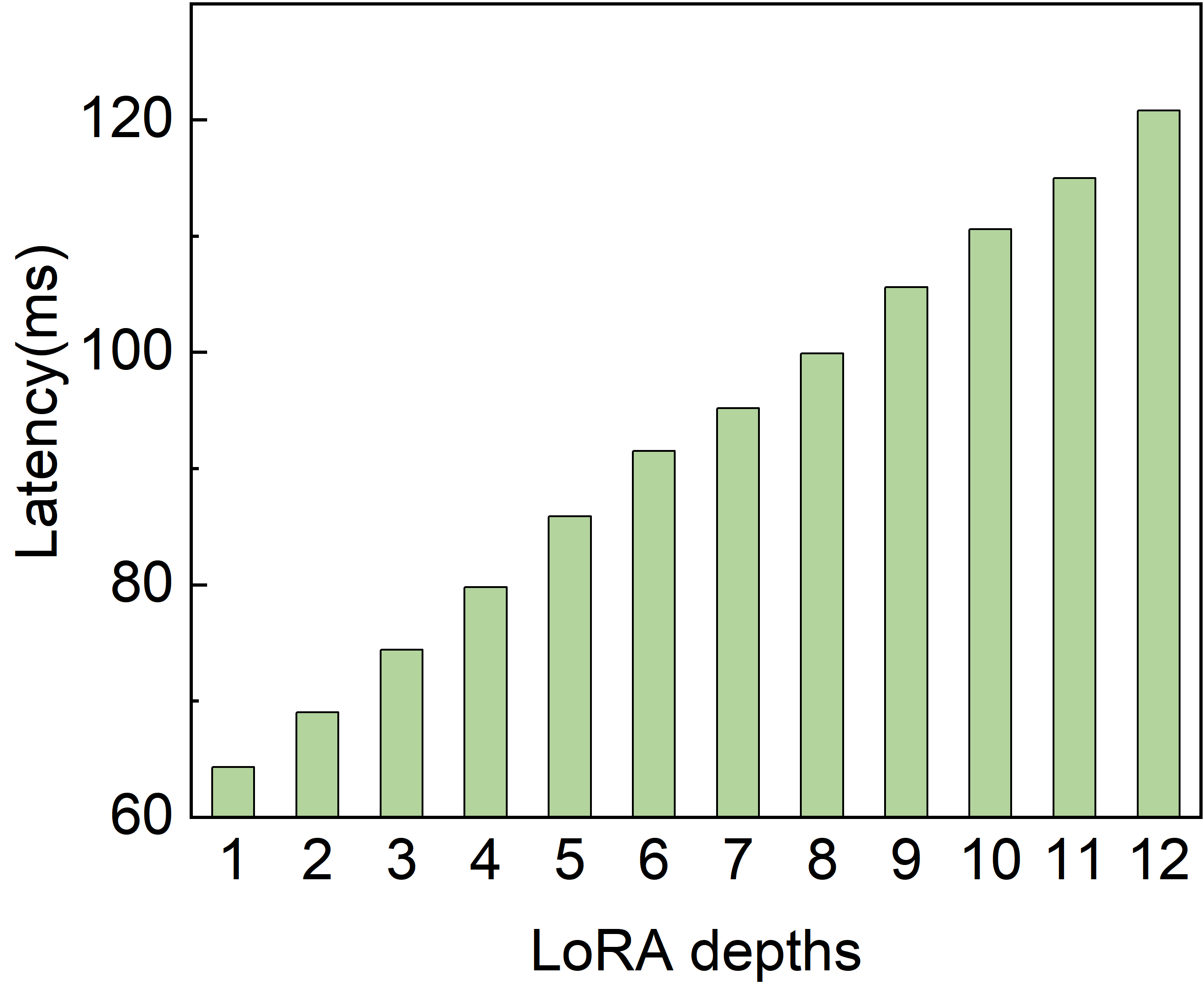}
        \end{minipage}%
        \label{3b}
    }%
    \centering
    \vspace{-0.5cm}
    \caption{ The impact of LoRA depths on fine-tuning.}
    \label{ob-3}
    \vspace{-0.6cm}
\end{figure}

The memory and computational overhead of LoRA is largely determined by the positions of the unfrozen layers. Specifically, updating the parameters of a given layer during training requires storing the activations of that layer and all subsequent layers. Moreover, to compute the gradient for that layer, the backward pass must propagate gradients from the output layer back to it. Our observations further reveal that both the position and number of unfrozen LoRA layers have a significant impact on fine-tuning performance, \ie, resource consumption and model accuracy. In particular, consecutively fine-tuning LoRA layers starting from the output demonstrates superior efficiency in both resource usage and accuracy improvement. With this approach, although increasing the number of trainable layers continues to improve performance, the marginal gains gradually diminish, while resource consumption grows nearly linearly. Based on these observations, we define the number of consecutively unfrozen LoRA layers starting from the output as the LoRA depth, which plays a critical role in balancing model accuracy and resource usage. To assess its impact, we conduct a series of experiments.

First, to verify the impact of the position of unfrozen LoRA layers on fine-tuning performance, we conduct a fine-tuning experiment using a 12-layer RoBERTa-base model~\cite{liu2019roberta} on the MNLI dataset~\cite{wang2018glue} (detailed experimental settings can be found in Section~\ref{evaluation}). In this experiment, we selectively fine-tune LoRA layers at different positions while freezing the remaining layers. Specifically, we evaluate four configurations: all layers fine-tuned (denoted as Layers-A), shallow layers \{0, 1, 2, 3\} (Layers-S), middle layers \{4, 5, 6, 7\} (Layers-M), and deep layers \{8, 9, 10, 11\} (Layers-D). The experimental results in Fig.~\ref{ob-2} lead to the following two key insights:

(1) Consecutive fine-tuning from the output layer is more resource-efficient. As shown in Fig.~\ref{2a}, despite fine-tuning only one third of the layers, Layers-S consumes 93\% of the memory used by Layers-A. In comparison, Layers-M and Layers-D consume 69.9\% and 46.5\% of the memory used by Layers-A, respectively. In LoRA fine-tuning, updating a given layer requires retaining the activations of all subsequent layers. As a result, although Layers-S fine-tunes only the first four layers, activations from all 12 layers must still be stored, leading to high memory usage.
Moreover, Layers-S exhibits $1.37\times$ and $1.87\times$ higher computation latency than Layers-M and Layers-D, respectively. This is because the trainable layers in Layers-S are located near the input, resulting in a longer backward pass and thus higher computational overhead.

(2) Shallow-layer fine-tuning provides limited performance gains. As shown in Fig.~\ref{2b}, Layers-M and Layers-D achieve accuracies of 77.7\% and 73.4\%, respectively, while Layers-S achieves only 64.1\%. This substantial gap indicates that tuning shallow layers has limited capacity to adapt to downstream tasks, whereas middle and deeper layers play a more important role in improving fine-tuning performance.

Therefore, fine-tuning consecutive LoRA layers from the output layer while freezing all other parameters is not only more resource-efficient but also straightforward and effective in preserving model accuracy. Then, we further investigate the relationship between LoRA depth and both resource usage and performance. We conduct a series of experiments using RoBERTa-base on MNLI, with LoRA depth ranging from 1 to 12. The experimental results, shown in Fig.~\ref{ob-3}, reveal the following two key findings:

(1) As LoRA depth increases, resource consumption exhibits an approximately linear growth trend. Specifically, each additional LoRA layer increases the computational latency by approximately 5 ms and memory usage by around 199 MB.

(2) Although fine-tuning performance improves with increasing LoRA depth, the performance gain gradually diminishes, particularly in the shallower layers. From Fig.~\ref{3a}, we observe that when LoRA depth increases from 1 to 8, model accuracy improves by 15.8\%, with an average accuracy gain of approximately 2.26\% per additional LoRA layer. However, when the depth increases from 8 to 12, the accuracy gain is only 0.4\%.

In summary, while performance continues to improve with increasing LoRA depth, the marginal performance gain progressively diminishes, while memory usage and computational latency grow in a near-linear fashion. Therefore, selecting an appropriate LoRA depth based on the capabilities of end devices is essential to balance fine-tuning performance and resource consumption.

\subsection{Motivations for Adaptive LoRA Depth and Activation Quantization}
\label{activation_quantization}

Previous observations indicate that stable fine-tuning performance during training typically requires a larger LoRA depth. However, increasing the LoRA depth may exceed the memory capacity of end devices, thereby making the deployment of LLMs infeasible in practical scenarios. 
During training, activation storage is the primary contributor to the memory bottleneck. Existing approaches to mitigate activation-related memory overhead generally fall into two categories, \ie, activation checkpointing and activation quantization. Activation checkpointing reduces memory usage by discarding intermediate activations during the forward pass and recomputing them during the backward pass as needed. While this approach significantly lowers memory consumption, it introduces considerable computational overhead. For instance, fine-tuning a RoBERTa-base model on the MNLI dataset using an A6000 GPU shows that enabling activation checkpointing increases per-batch latency by approximately 83\%.
In contrast, activation quantization reduces memory usage by quantizing activations during the forward pass and dequantizing them during backpropagation for gradient computation. Although this method also incurs extra computation, we optimize the quantization and dequantization operations using Triton~\cite{tillet2019triton} as in Jetfire~\cite{xi2024jetfire} (see Section~\ref{implementation} for details), resulting in only a 36\% increase in per-batch latency. Additionally, we implement layer-wise quantization to better balance computational overhead.

\begin{figure}[t]
    \centering
    \subfigure[ Accuracy \& Memory usage]{
        \begin{minipage}[t]{0.5\linewidth}
        \centering
        \includegraphics[width=1.7in]{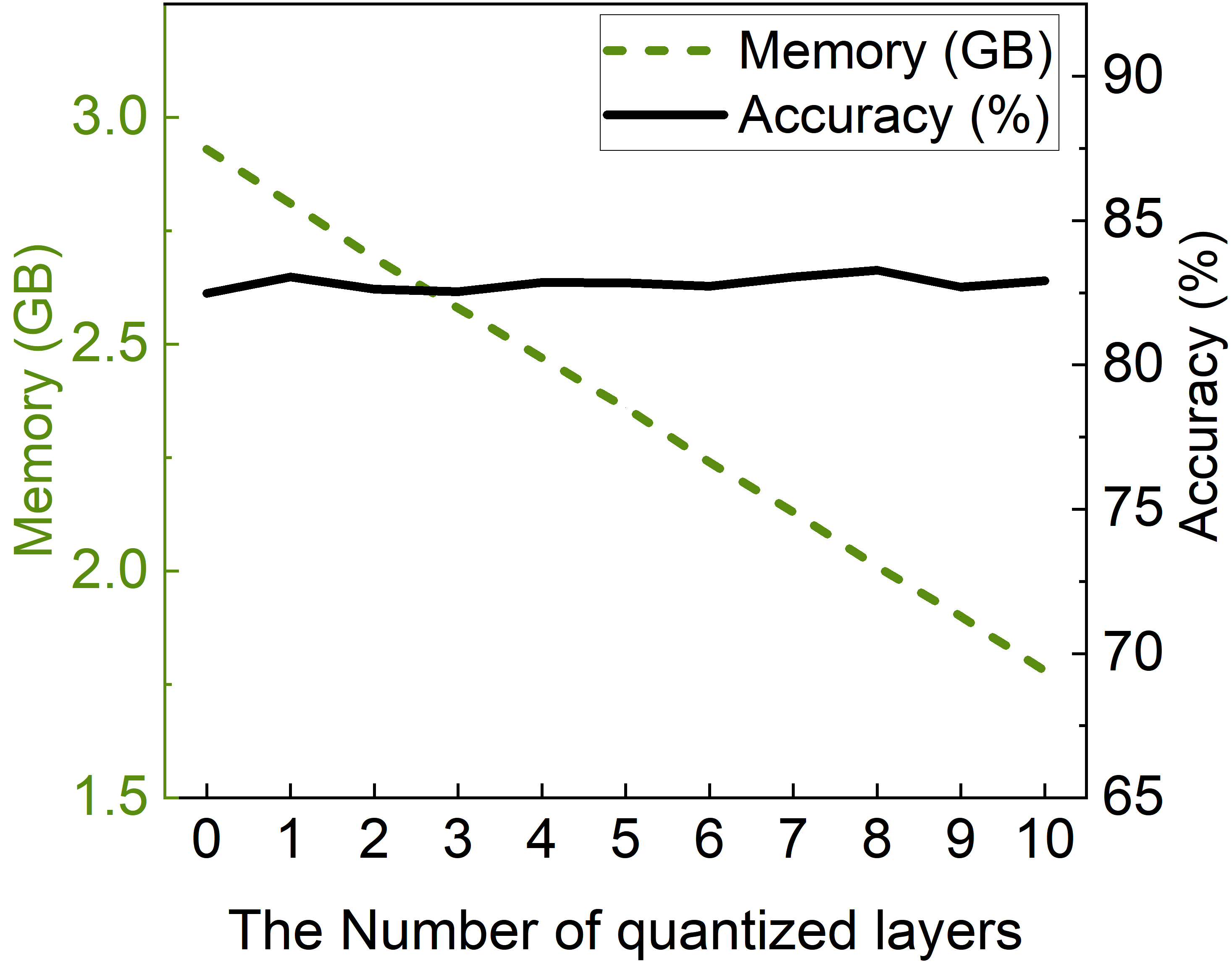}
        \end{minipage}%
        \label{4a}
    }%
    \subfigure[Accuracy \& Latency]{
        \begin{minipage}[t]{0.5\linewidth}
        \centering
        \includegraphics[width=1.7in]{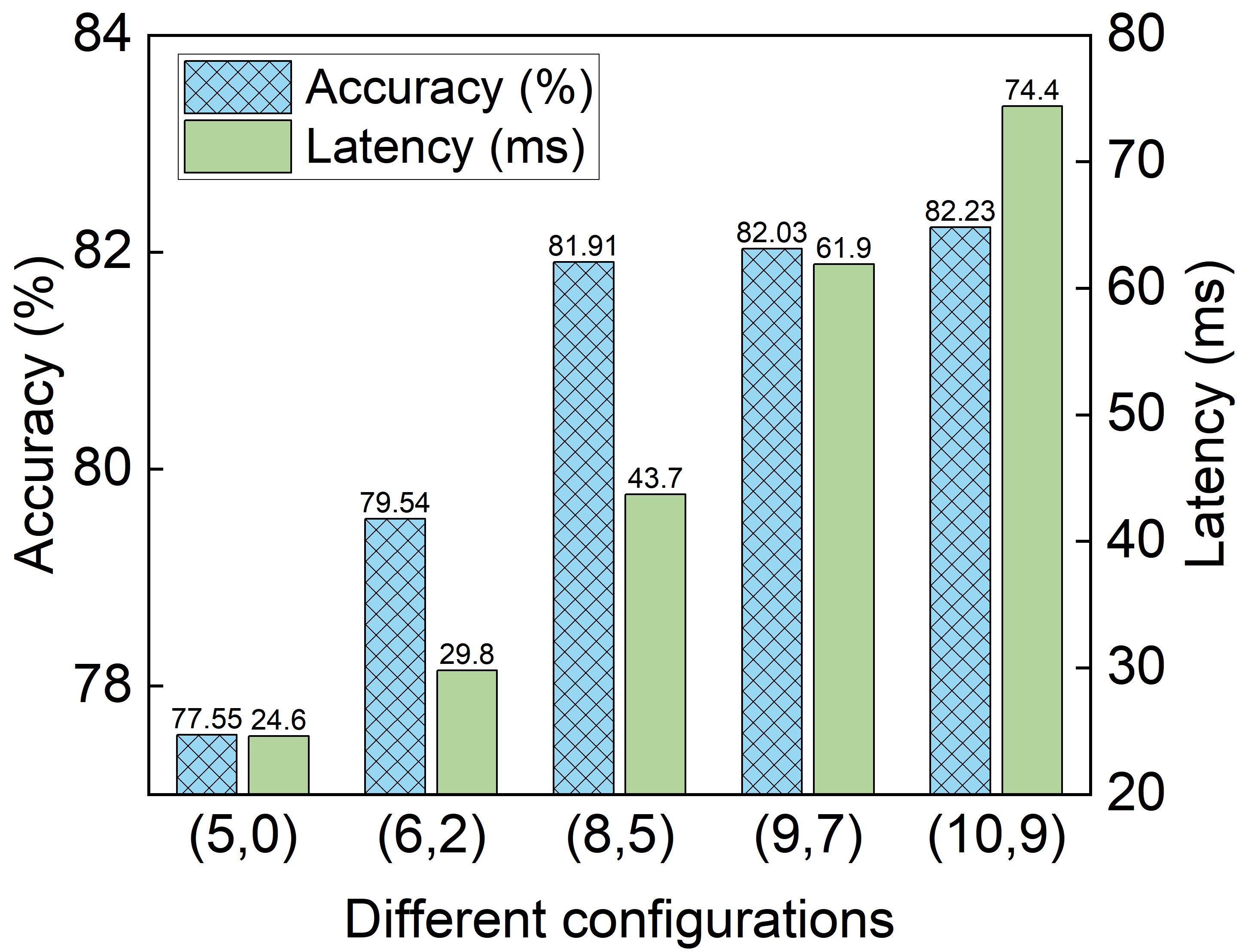}
        \end{minipage}%
        \label{4b}
    }%
    \centering
    \vspace{-0.5cm}
    \caption{The impact of activation quantization on fine-tuning.}
    \label{ob-4}
    \vspace{-0.6cm}
\end{figure}

We define the number of activation quantization layers as the number of consecutive transformer layers whose activations are quantized during training, starting from the first unfrozen LoRA layer. Taking RoBERTa-base as an example, as shown in Fig.~\ref{4a}, we enable all LoRA layers for fine-tuning and apply activation quantization sequentially from the first transformer layer onward. This approach results in an average memory reduction of 58\% when fine-tuning the corresponding LoRA layers. Moreover, when the number of quantized layers reaches 8, the model achieves a 0.8\% accuracy improvement compared to the non-quantized baseline. This performance gain is attributed to the stochastic noise introduced during quantization, which helps prevent overfitting and enhances the model’s generalization capability.

To investigate the synergy between the number of activation quantization layers  and LoRA depth under fixed memory constraints, we systematically vary the configuration, represented as (LoRA depth, number of quantized layers), and evaluate the resulting performance. For example, a configuration of (8, 5) represents a LoRA depth of 8 and the number of quantized layers of 5. As shown in Fig.~\ref{4b}, reallocating memory saved through activation quantization to support larger LoRA depth can substantially improve model accuracy. For instance, the configuration (5, 0) achieves an accuracy of 77.55\%, while (8, 5) reaches 81.91\%, yielding an absolute improvement of 4.36\%. However, as both LoRA depth and the number of quantized layers continue to increase, the marginal accuracy gain gradually diminishes. For example, increasing from (8, 5) to (10, 9) improves accuracy by only 0.32\%, indicating a clear saturation effect. Meanwhile, computational latency consistently increases. When moving from  (8, 5) to (10, 9), the delay rises from 43.7ms to 74.4ms, representing a 70.3\% increase.

While increasing the number of quantized layers can effectively support larger LoRA depth and enhance fine-tuning performance, the resulting increase in latency may cause slower devices to become stragglers, thereby prolonging synchronization time during training.
Therefore, to accelerate overall convergence, it is necessary to explore the optimal configuration of the number of activation quantization layers and LoRA depth according to the resource constraints of heterogeneous devices.



\section{System Design}\label{sec:algorithm}
\subsection{System Overview}
\begin{figure}[t]
    \centering
    \includegraphics[width=1\columnwidth]{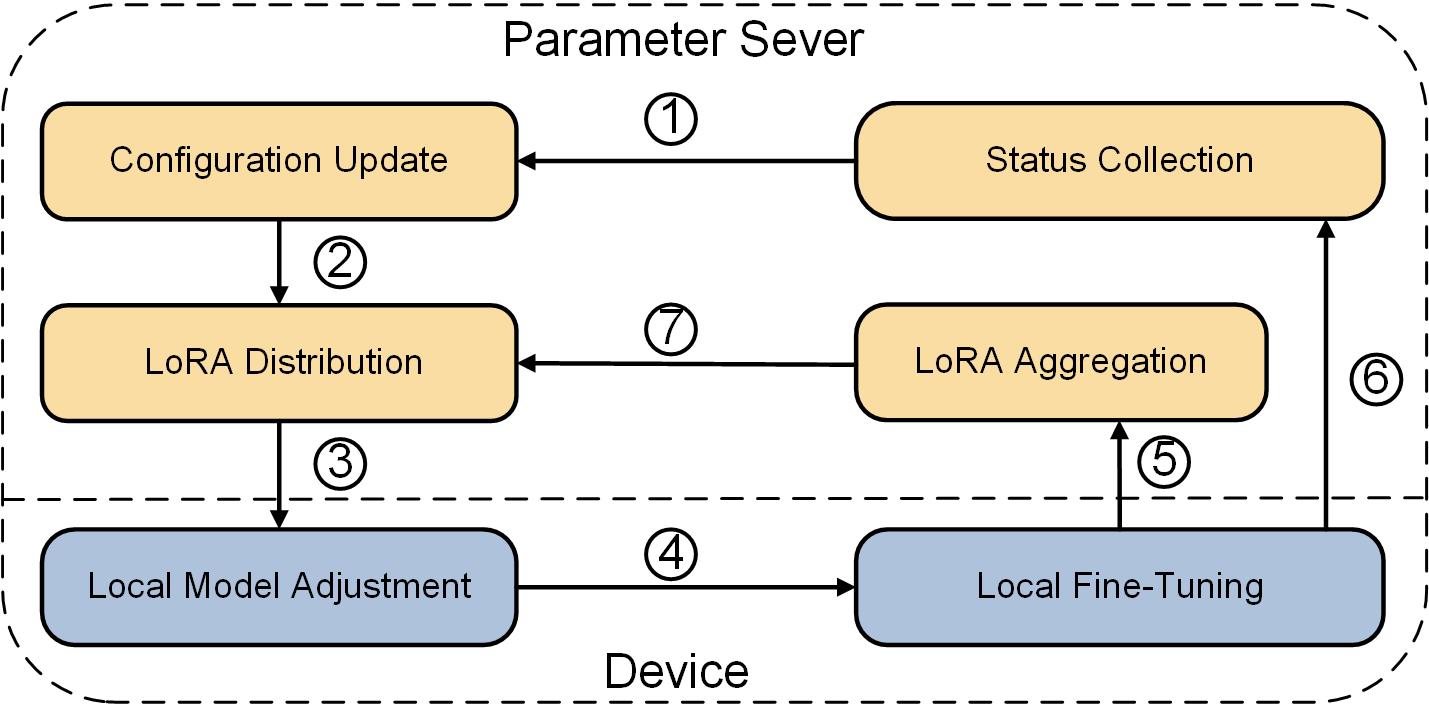}
    \vspace{-0.5cm}
    \caption{Overview of FedQuad’s workflow.}
    \label{framework}
\vspace{-0.3cm}
\end{figure}

Inspired by the above observations, we propose FedQuad, a novel and efficient federated fine-tuning framework tailored for end devices with constrained resource and heterogeneous system characteristics. Specifically, FedQuad determines the optimal fine-tuning configurations for each device based on their status information. Upon completion of local fine-tuning, FedQuad performs adaptive aggregation of the updated parameters and subsequently decides the optimal configurations for the next training round. The overall workflow of FedQuad comprises six main steps (as illustrated in Fig. ~\ref{framework}):

\textbf{Status Collection.} At the beginning of each training round, the parameter server (PS) collects status information from each device (\normalsize{\textcircled{\scriptsize{1}}}), including available memory and computational capabilities, to inform the configuration updates.

\textbf{Configuration Update.} The PS updates appropriate configurations, including LoRA depth and the number of activation quantization layers to each device based on their current memory availability and computational capabilities (\normalsize{\textcircled{\scriptsize{2}}}). This ensures efficient resource utilization while simultaneously avoiding out-of-memory errors and degraded fine-tuning efficiency.

\textbf{LoRA Distribution.} The PS disseminates the aggregated LoRA weights along with the updated configurations to each device (\normalsize{\textcircled{\scriptsize{3}}}).

\textbf{Local Model Adjustment.} 
Upon receiving its corresponding configuration, each device reconfigures its local model by adjusting unfrozen LoRA layers and quantized activation layers. After adjusting the settings of individual layers within the model, each device updates its local LoRA parameters to align with the received global parameters before starting the local fine-tuning phase (\normalsize{\textcircled{\scriptsize{4}}}).

\textbf{Local Fine-Tuning.} End devices perform local fine-tuning while monitoring runtime performance such as memory utilization and computational latency. After completing the local training, devices upload the updated LoRA parameters (\normalsize{\textcircled{\scriptsize{5}}}) and recorded performance status (\normalsize{\textcircled{\scriptsize{6}}}) to the PS .

\textbf{LoRA Aggregation.} The PS performs adaptive weighted aggregation of the received LoRA parameters, considering the varying LoRA depths used by individual devices, to obtain the updated global model (\normalsize{\textcircled{\scriptsize{7}}}).

The overall procedure iterates over multiple training rounds until the model satisfies a predefined convergence condition or achieves the target accuracy.

\subsection{Status Collection}
FedQuad effectively addresses resource constraints and system heterogeneity by allocating optimal training configurations to each device. 
Specifically, in the $h$-th training round, FedQuad determines a tailored configuration $(d_i^h, a_i^h)$ for each device $i$ according to its resource constrains, where $d_i^h$ represents the assigned LoRA depth and $a_i^h$ represents the number of activation quantization layers.

Before determining these configurations, each device uploads its current status, including available memory, computational capacity.
The available memory of device $i$ in round $h$ is denoted by $M_i^h$, representing the portion of GPU or system memory that can be safely used for training without interfering with other local applications.

In round $h$, the computational capability $q_i^h$ denotes the floating-point processing throughput of device $i$. Accordingly, the local training time $\mu_i^h$ incurred by end device $i$ during round $h$ is calculated as:
\begin{equation}
u_i^h = \frac{C(d_i^h, a_i^h)}{q_i^h}
\end{equation}
where $C(\cdot)$ denotes the computational complexity of the task, modeled as a linear function of LoRA depth $d$ and the number of quantized layers $a$.

\subsection{Configuration Update}

Based on each device status reports, FedQuad constructs a unified model that incorporates accuracy, memory, and latency to guide configuration decisions.
As discussed in Section~\ref{sec:motivation}, fine-tuning performance is primarily determined by the LoRA depth $d$. 
Let $D^h$ denote the set of LoRA depths assigned to all devices in the $h$-th training round, and let $D$ be the union of all such depths across $H$ training rounds, where $H$ denotes the total number of training rounds:

\begin{equation}
D^h = \{ d_i^h \mid i \in [1, n] \}
\end{equation}

\begin{equation}
D = \bigcup_{h=1}^{H} D^h = \bigcup_{h=1}^{H} \left\{ d_i^h \mid i \in [1, n] \right\}
\end{equation}

Let $\Delta w^H(D)$ represent the aggregated global LoRA parameters after $H$ training rounds under the set $D$. The loss function of the global model after training can be expressed as $F(w_0, \Delta w^H(D))$. The convergence requirement after training can be formulated as the following constraint:
\begin{equation}
\label{lim_1}
F(w_0, \Delta w^H(D)) - F(w_0, \Delta w^*) \leq \epsilon
\end{equation}
where $\epsilon > 0$ denotes the convergence threshold, and $F(w_0, \Delta w^*)$ represents the optimal fine-tuning performance. Ideally, $\epsilon$ should approach zero, indicating that the trained model closely approximates the optimal solution~\cite{xu2022adaptive}.

On resource-constrained devices, the number of LoRA layers that can be fine-tuned is typically limited. However, the memory saved through activation quantization can be reallocated to support a larger LoRA depth, enabling the fine-tuning of more layers and thereby improving fine-tuning accuracy.
Therefore, let $m_o$ denote the additional memory overhead incurred by increasing the LoRA depth by one layer, $m_q$ denote the memory savings achieved by quantizing the activations of a single transformer layer, and $m_f$ denote the fixed memory overhead during training (\eg, model parameters). Then, the memory constraint can be formulated as:
\begin{equation}
\label{lim_2}
m_f + m_o \cdot d_i^h - m_q \cdot a_i^h \leq M_i^h, \quad \forall i \in [1, n], \forall h \in [1, H]
\end{equation}

The completion time $t_i^h$ consumed by device $i$ in round $h$ includes both local training time and communication time. Nevertheless, given the significantly small size of LoRA parameters, the communication time is omitted in subsequent analysis \cite{wu2023fedlora, zhang2023fedpetuning}.
Therefore, the completion time can be approximated as:

\begin{equation}
\label{completion time}
t_i^h = u_i^h 
\end{equation}

Strong devices have to wait for weak devices to complete training 
in each round. Let $t_h$ denote the longest completion time across all participating devices in round $h$. We estimate the average waiting time across all devices in round $h$ as: 
\begin{equation} W_h = \frac{1}{n} \sum_{i=1}^{n} (t_h - t_i^h) \end{equation}
A larger $W_h$ indicates greater synchronization delay, which can slow down overall convergence. To mitigate this issue, we impose a constraint on the average waiting time in each round. Specifically, we require that:

\begin{equation}
\label{lim_4}
W_h = \frac{1}{n} \sum_{i=1}^{n} (t_h - t_i^h) \leq \theta, \quad \forall i \in [1, n], \forall h \in [1, H]
\end{equation}
where $\theta$ is a predefined threshold that limits the allowable average waiting time.

As both LoRA layer freezing and activation quantization are applied on a per-layer basis, the LoRA depth $d$ and the number of quantized layers $a$ must be non-negative integers. Specifically, $d$ has to satisfy $d \in [1, L]$, where $L$ denotes the total number of transformer layers, and $a$ should satisfy $a \in [0, d-1]$, since we notice that quantizing the final output layer tends to degrade fine-tuning accuracy while incurring additional memory overhead. Thus, the constraints can be formulated as: 
\begin{equation} 
\label{range-lim} d, a \in \mathbb{Z}_{\geq 0}, \quad d \in [1, L], \quad a \in [0, d-1] \end{equation}
where $\mathbb{Z}_{\geq 0}$ denotes the set of non-negative integers.

Given a FedFT task, FedQuad aims to assign an appropriate configuration $(d_i^h, a_i^h)$ to each device $i$ in each round $h$, based on its available resources, to achieve the target accuracy while minimizing the total fine-tuning time $\sum_{h=1}^{H} t_h$. Thus, we can formulate the problem as: 
\begin{equation}\label{problem}
\begin{array}{c}
\min \quad \sum_{h=1}^{H} t_h \\
\text{s.t.} \quad \eqref{lim_1},\ \eqref{lim_2},\ \eqref{lim_4},\ \eqref{range-lim}
\end{array}
\end{equation}


\begin{algorithm}[t]
\caption{Adaptive Configuration Selection (ACS) in FedQuad}
\label{alg}
\KwIn{
    Device memory constraint $M_i^h$ at round $h$;\\
    Compute capacity $u_i^{h{-}1}$ at round $h$;\\
    Average completion time $t_{\text{avg}}^{h{-}1}$ in round $h{-}1$;
}
\KwOut{Optimal configuration $(d_i^h, a_i^h)$ for each device $i$ at round $h$}

\textcolor{violet}{\tcp{Step 1:  Identify feasible and efficient configurations $(d, a)$ under memory constraint}} \label{alg:step1-begin}
\For{each device $i \in [1, n]$}{
    $\mathcal{C}_i \gets \emptyset$; \\
    $a_{\text{cur}} \gets 0$; \\
    \For{$d \in [1, d^{\max}]$}{
        \For{$a \in [a_{\text{cur}}, d{-}1]$}{
            \If{the memory constraint in Eq.~\eqref{lim_2} holds}{
                Add $(d, a)$ to $\mathcal{C}_i$; \\
                $a_{\text{cur}} \gets a$; \\
                \textbf{break} \tcp*{Once a valid pair $(d, a)$ is found, break out of the inner loop}
            }
        }
    }
} \label{alg:step1-end}

\textcolor{violet}{\tcp{Step 2: Estimate completion time for each feasible configuration}} \label{alg:step2-begin}
\For{each device $i \in [1, n]$}{
    Estimate $t_i^h$ for each $(d, a) \in \mathcal{C}_i$ using $\mu_i^{h{-}1}, b_i^{h{-}1}$ (see Eq.~(\ref{completion time}));
} \label{alg:step2-end}

\textcolor{violet}{\tcp{Step 3: Compute performance gain based on gradient norms}} \label{alg:step3-begin}
\For{each LoRA depth $d \in [1, L]$}{
    Compute $G(d) = \sum_{l = L{-}d}^{L{-}1} g_l$;
} \label{alg:step3-end}

\textcolor{violet}{\tcp{Step 4: Select the configuration that maximizes reward}} \label{alg:step4-begin}
\For{each device $i \in [1, n]$}{
    $(d_i^h, a_i^h) \gets \displaystyle\mathop{\arg\max}_{(d, a) \in \mathcal{C}_i} R(d, a)$;
} \label{alg:step4-end}

\Return $\{(d_i^h, a_i^h) \mid i \in [1, n]\}$;
\end{algorithm}

To solve the problem in Eq. (\ref{problem}), we adopt a greedy strategy ACS, as shown in Algorithm\ref{alg}, to determine the optimal configuration for each device at every round. ACS first identifies the feasible and efficient configurations $(d, a)$ for each device based on its memory constraint (Lines \ref{alg:step1-begin}– \ref{alg:step1-end}). Specifically, the PS iterates through all possible LoRA depths $d$ for each device $i$ according to its available memory $M_i^h$, and adjusts the number of activation quantization layers $a$ accordingly. 
For each candidate depth $d$, it identifies the minimal number of quantized layers $a$  that satisfies the memory constraint in Eq.~\eqref{lim_2} while avoiding additional computational overhead. The corresponding configuration $(d, a)$ is then added to the feasible set $\mathcal{C}_i^h$.

Next, FedQuad evaluates the training cost and performance gain of each candidate configuration (Lines \ref{alg:step2-begin}– \ref{alg:step3-end}). Based on the current device’s computational capabilities, FedQuad estimates the completion time $t_i^h$ for every feasible configuration $(d, a)$. Additionally, FedQuad records the average completion time $t^{h-1}_{\text{avg}}$ from the previous round as an estimate for the current round’s average completion time. FedQuad aims to minimize the difference between each device’s completion time and $t^{h-1}_{\text{avg}}$, enabling strong devices to fully utilize their computational resources while ensuring weak devices do not cause excessive synchronization delays.

Since gradient information inherently captures the marginal contribution of each layer to the loss function \cite{sun2024exploring, he2025gora}, FedQuad quantifies the performance gain $G(d)$ associated with a LoRA depth of $d$ based on layer-wise gradient norms. Specifically, FedQuad computes the gradient norm $g_l$ for each LoRA layer $l \in [0, L{-}1]$ in the global model and aggregates the norms of the top $d$ layers. Formally, the performance gain is defined as: \begin{equation} G(d) = \sum_{l = L - d}^{L - 1} g_l \end{equation}

Finally, to balance performance gain and training cost, FedQuad defines a reward function to evaluate the trade-off of each configuration $(d, a)$ for each device $i$:
\begin{equation}
\label{reward}
R(d, a) = \frac{G(d)}{t_i^h(d, a) - t^{h-1}_{\text{avg}} + c}
\end{equation}
where the numerator $G(d)$ represents the contribution of each configuration to model convergence. The denominator captures the gap between the completion time under the current configuration $(d, a)$ and the average completion time. A smaller gap indicates that the selected configuration better matches the device's computational capability, resulting in a higher reward. Therefore, a higher $R(d, a)$ reflects a more favorable trade-off between accuracy and efficiency. The pair $(d, a)$ achieving the highest $R(d, a)$ is chosen as the optimal configuration for device $i$ (Lines \ref{alg:step4-begin}– \ref{alg:step4-end}).

\begin{figure*}[t]
    \centering
    \includegraphics[width=\textwidth]{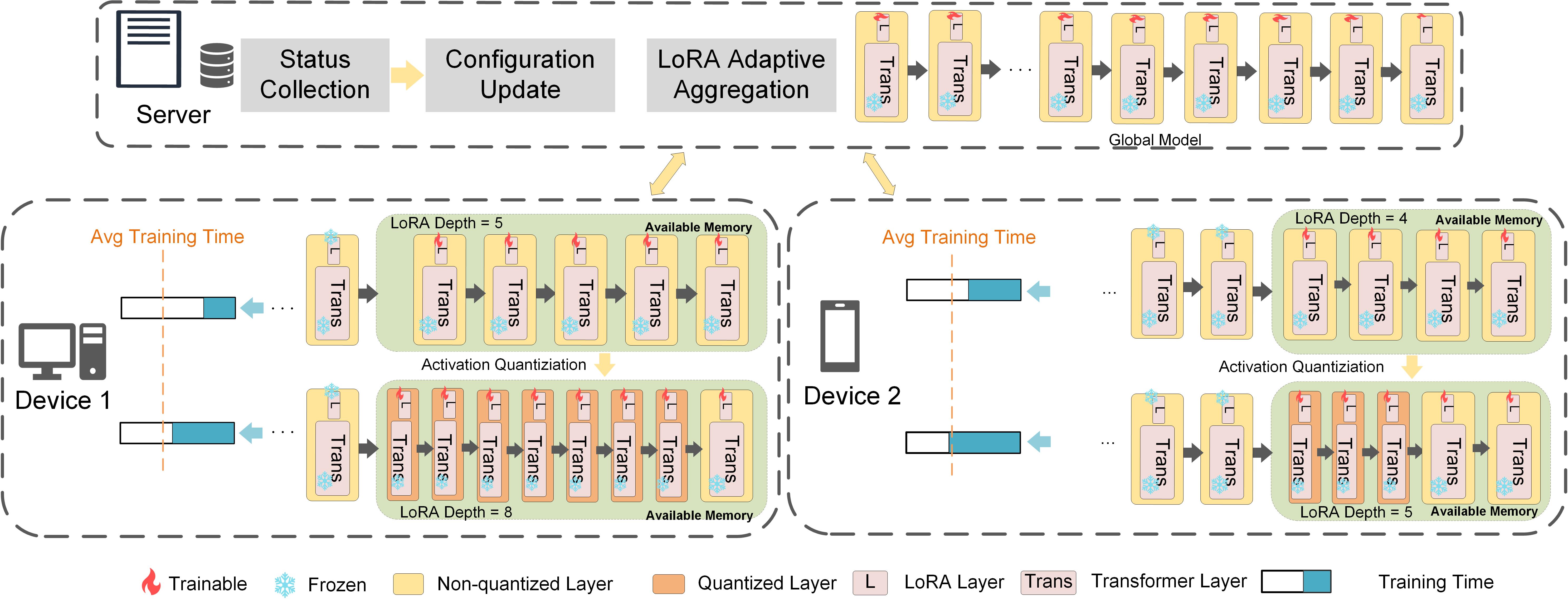}
    \vspace{-0.6cm}
    \caption{An illustrative example of FedQuad.}
    \label{illu}
    \vspace{-0.4cm}
\end{figure*}

\subsection{Local Model Adjustment}

In round $h$, device $i$ receives the LoRA parameters $\Delta \omega^h$ and the training configuration $(d_i^h, a_i^h)$ from PS. Based on this configuration, the device first adjusts the trainable LoRA layers, represented by $L_t$, as:  
\[
L_t = \{ l \,|\, l \in [L - d_i^h, L - 1] \},  \forall i \in [1, n], \, \forall h \in [1, H].
\]  
The LoRA layers within $L_t$ are set to be trainable, while all other layers remain frozen.
Subsequently, the device configures the activation quantization transformer layers, denoted as $L_q$, for the current round as follows:  
\[
L_q = \{ l \,|\, l \in [L - d_i^h, L - d_i^h + a_i^h - 1] \}, \forall i \in [1, n], \, \forall h \in [1, H].
\]  
Once the fine-tuning configuration adjustments are complete, the device incorporates updates received from the PS to align its local model with the global model. Specifically, for a LLM with pre-trained parameters $\omega_0 = \{\omega_l^0 \,|\, l \in [0, L-1]\}$,
 the device $i$ updates its local model as:  
\[
\omega_i^h = \{\Delta \omega^h, \omega_0\}.
\]  
After local model adjustment, device $i$ trains the model on the local dataset $\mathcal{D}_i$.
\subsection{LoRA Aggregation}

Upon receipt of the updated LoRA parameters from all devices, the PS performs  global aggregation. Because devices may fine-tune differing numbers of LoRA layers, not every layer update is present on every device. To address this heterogeneity, the PS employs an adaptive, layer-wise aggregation protocol. Specifically, the aggregation process for each layer only considers the updates from the devices that have valid updates for that particular layer, thereby enhancing the performance of the global model for that layer.

The PS aggregates the available updates layer by layer. Let the global LoRA update for layer $l$ be denoted as $\Delta\boldsymbol{\omega}_l^{h+1}$, which is obtained by aggregating the updates from $n_l$ devices. The adaptive layer-wise aggregation rule is defined as:

\begin{equation} 
\Delta\boldsymbol{\omega}_l^{h+1} = \frac{1}{n_l} \sum_{i=1}^{n_l} \Delta\boldsymbol{\omega}_{i,l}^h
\end{equation}
where $\Delta\boldsymbol{\omega}_{i,l}^h$ denotes the update to the LoRA layer $l$ by device $i$ during the round $h$. $n_l$ represents the number of devices with valid updates for layer $l$. $\Delta\boldsymbol{\omega}_l^{h+1}$ represents the aggregated global update for layer $l$.

After aggregating the LoRA parameters, FedQuad initiates a new round of status collection to support subsequent configuration updates, thereby guiding the fine-tuning process on each device based on its latest resource constrains. 
By adaptively assigning larger LoRA depths and more quantized layers to stronger devices, and smaller configurations to weaker ones based on their resource limitations, FedQuad improves fine-tuning performance and efficiency. Consequently, it effectively addresses the dual challenges of resource constraints and system heterogeneity.

\subsection{An Illustrative Example of FedQuad}

As illustrated in Fig.~\ref{illu}, FedQuad determines the optimal configuration for each device based on its resource constraints to accelerate overall convergence. For strong devices, although their greater memory capacity allows for larger LoRA depth during fine-tuning, the prolonged completion times of weaker devices often result in extended  idle waiting for the strong devices. To mitigate this, FedQuad increases the number of activation quantization layers to support larger LoRA depths, thereby fully utilizing the computational resources of strong devices and improving fine-tuning performance. For example, as shown for Device 1 in Fig.~\ref{illu}, its available memory allows only limited LoRA layers, resulting in a relatively smaller contribution to model convergence and a much shorter completion time than the average, which indicates underutilized computing capacity. FedQuad increases its quantization layers to support a larger LoRA depth. Although this leads to a slight increase in training time, Device 1 still completes earlier than average, thus causing no synchronization delays and helping reduce overall idle time.
For weaker devices, constrained in both memory and computational power, aggressively increasing quantization layers to enable larger LoRA depth may improve accuracy but risks introducing excessive synchronization delay. Therefore, FedQuad adjusts both activation quantization and LoRA depth, aiming to improve fine-tuning performance without overburdening the device. This prevents the device from becoming a stragglers. For instance, FedQuad assigns Device 2 additional quantization layers to enhance fine-tuning performance, but limits the number to ensure that its completion time remains close to the average, avoiding significant slowdowns to the overall training process.
After all end devices complete local training, they upload their parameter updates and status information to the PS. Upon receiving these updates, the PS performs adaptive aggregation of the LoRA parameters, while simultaneously leveraging the devices' status information to determine the optimal configurations for the subsequent round.


\section{Performance Evaluation}\label{sec:evaluation}

\subsection{Experimental Settings}
\label{evaluation}
\textbf{System Implementation.}
\label{implementation}
We implement the FedQuad prototype based on the open-source FedPETuning framework~\cite{zhang2023fedpetuning}, enhancing its functionality with approximately 2,500 lines of custom code. To ensure compatibility with modern LLM architectures, FedQuad is seamlessly integrated with the widely used \textit{Transformers} library~\cite{huggingface2024transformers}, and leverages its modular APIs to support LLM initialization and adaptation.
To enable activation quantization, we modify the core LLM modules in the transformers.models package. Following the observations reported in SLIMFIT~\cite{ardakani2023slimfit}, we note that static activations from nonlinear operations such as GELU, MatMul, Softmax, and LayerNorm cannot be entirely eliminated by freezing the associated layers. Therefore, during the forward pass, we quantize the activations of these functions and subsequently apply dequantization during the backward pass. This approach effectively reduces memory overhead without compromising model accuracy. 
Our quantization process follows the approach proposed in Jetfire~\cite{xi2024jetfire}, and is implemented using Triton~\cite{tillet2019triton}, a domain-specific programming language and compiler designed for developing efficient deep learning primitives. In our experiments, we set the block size to 32, \ie, $B=32$.
We adopt PyTorch~\cite{paszke2019pytorch} as the underlying training framework to manage the local fine-tuning workflow. GPU acceleration is supported by CUDA v12.3 and cuDNN v9.1.0. For distributed communication between edge devices and the central server, we utilize torch.distributed, a native PyTorch library for parallel and distributed training.

\textbf{Models.} 
We primarily evaluate FedQuad on three popular LLMs: BERT-large~\cite{devlin2018bert} with 336M parameters, RoBERTa-large~\cite{liu2019roberta} with 355M parameters, and DeBERTaV3-large~\cite{he2021debertav3} with 435M parameters. These models have been extensively adopted in prior federated fine-tuning studies~\cite{lin2021fednlp, cai2022fedadapter, zhang2023fedpetuning, liu2022no}. Each model consists of 24 transformer layers. Pre-trained weights are obtained from the Hugging Face repository~\cite{huggingface2025models}. 

\textbf{Datasets.} 
We conduct our experiments on four widely used NLP datasets from the GLUE benchmark \cite{wang2018glue} (summarized in Table \ref{tab-glue-datasets}):
(1) The Multi-Genre Natural Language Inference (MNLI) dataset is a large-scale dataset containing over 400,000 sentence pairs used for training and evaluating LLMs on natural language inference (NLI) tasks. In the NLI tasks, the goal is to determine whether a premise entails, contradicts, or is neutral with respect to a given hypothesis. 
(2) The Quora Question Pairs (QQP) dataset consists of more than 400,000 question pairs collected from the Quora platform. Each pair is labeled to indicate whether the two questions are paraphrases of each other or not. QQP is commonly utilized for training and evaluating LLMs on paraphrase detection tasks. (3) The Question Natural Language Inference (QNLI) dataset is derived from the Stanford Question Answering Dataset (SQuAD). It consists of over 100,000 sentence–question pairs labeled to indicate whether the sentence contains the correct answer to the corresponding question. 
(4) The Stanford Sentiment Treebank Binary Version (SST-2) dataset contains 70,042 movie review sentences annotated with binary sentiment labels. Each sentence is labeled as either positive or negative. 
To simulate non-independent and identically distributed (non-i.i.d.) data across devices, we follow prior work \cite{lin2021fednlp, cai2022fedadapter, zhang2023fedpetuning} by partitioning the dataset using a Dirichlet distribution. Specifically, for each device, we sample training data according to $D \sim \text{Dir}(\alpha)$, where $\alpha$ controls the degree of non-i.i.d. Unless otherwise specified, we set $\alpha = 10$ throughout all experiments.

\begin{table}[!t]
    \caption{Technical specifications of Jetson kits.}
    \vspace{-0.3cm}
    \centering
    \begin{tabular}{|l|c|c|}
        \hline
        & \textbf{AI Performance} & \textbf{GPU Type} \\ \hline
        Jetson TX2 & 1.33 TFLOPS & 256-core Pascal \\ \hline
        Jetson NX & 21 TOPS & 384-core Volta\\ \hline
        AGX Xavier & 32 TOPS &   512-core Volta \\ \hline \hline
        & \textbf{CPU Frequency} & \textbf{GPU Frequency} \\ \hline 
        Jetson TX2 & 1.2GHz & 0.85Ghz\\ \hline
        Jetson NX & 1.2GHz & 0.8Ghz\\ \hline
        Jetson AGX & 1.45GHz & 0.9Ghz\\ \hline
    \end{tabular}
    \label{jetson-info}
  
\end{table}

\begin{table}
\footnotesize
    \caption{Datasets and numbers of samples used in the experiments.}
    \vspace{-0.3cm}
    \centering
    \resizebox{\linewidth}{!}{
    \begin{tabular}{|c|c|c|c|}
    \hline
    Dataset & Application  & \# Train & \# Test  \\
    \hline
    MNLI     & Textual Entailment    & 392,702   & 9,815     \\ \hline
    QQP      & Semantic Equivalence  & 363,846   & 40,430    \\ \hline
    QNLI     & Question Answering    & 104,743   & 5,463     \\ \hline
    SST-2    & Sentiment Analysis    & 67,349    & 1,821     \\ \hline

    \end{tabular}}
    \label{tab-glue-datasets}
    \vspace{-0.5cm}
\end{table}
\textbf{Hardware.} 
Consistent with previous federated learning literature \cite{cai2022fedadapter, zhang2023fedpetuning, xu2024fwdllm}, our experiments are conducted in a semi-simulated manner using an AMAX deep-learning workstation equipped with eight NVIDIA A6000 GPUs. On-device training times are measured on three representative edge devices (summarized in Table \ref{jetson-info}):
(1) Jetson TX2 : A compact embedded computing platform designed specifically for AI applications at the edge.
(2) Jetson NX : Capable of accelerating computation by up to 21 TOPS, offering sufficient parallel computing power for running LLMs.
(3) Jetson AGX : The most powerful among the three, delivering computational capabilities of up to 32 TOPS.
Both TX2 and NX support four computational modes, whereas AGX supports eight modes. Devices operating in different modes demonstrate distinct performance levels.

\textbf{System Setup.} 
Our experiments involve a total of 100 devices, categorized into three types (strong, moderate, and weak) based on computational performance and memory capacity. To simulate the resource constraints of heterogeneous end devices, we adopt the following experimental setup:

1)\textbf{For Memory.} 
For clarity, we use the tunable LoRA depth in FedLoRA (i.e., federated fine-tuning with vanilla LoRA) to represent each device’s available memory capacity. For all methods involved, we convert the LoRA depth to the corresponding memory capacity to ensure a fair comparison. Furthermore, we assign dynamic depth ranges to different device categories: strong devices are assigned LoRA depths in the range of 18–24, moderate devices in the range of 11–17, and weak devices in the range of 4–10. To further simulate fluctuations in memory availability under real-world conditions, we randomly adjust each device’s tunable LoRA depth within its respective range after each training round. 

2)\textbf{For Compute.} 
We map the strong, moderate, and weak device categories to NVIDIA Jetson AGX, Jetson NX, and Jetson TX2 devices, respectively, representing varying levels of computational resources levels. Each Jetson device supports multiple operating modes corresponding to different computational capabilities and power configurations. We construct diverse computational profiles by setting various operation modes \cite{liu2024enhancing}. To realistically emulate fluctuations in computational resources, each device randomly switches its operating mode every 10 training rounds. 

\begin{figure*}
	\centering
	\subfigure[MNLI]{
		\begin{minipage}[b]{0.23\textwidth}
			\includegraphics[width=1\textwidth]{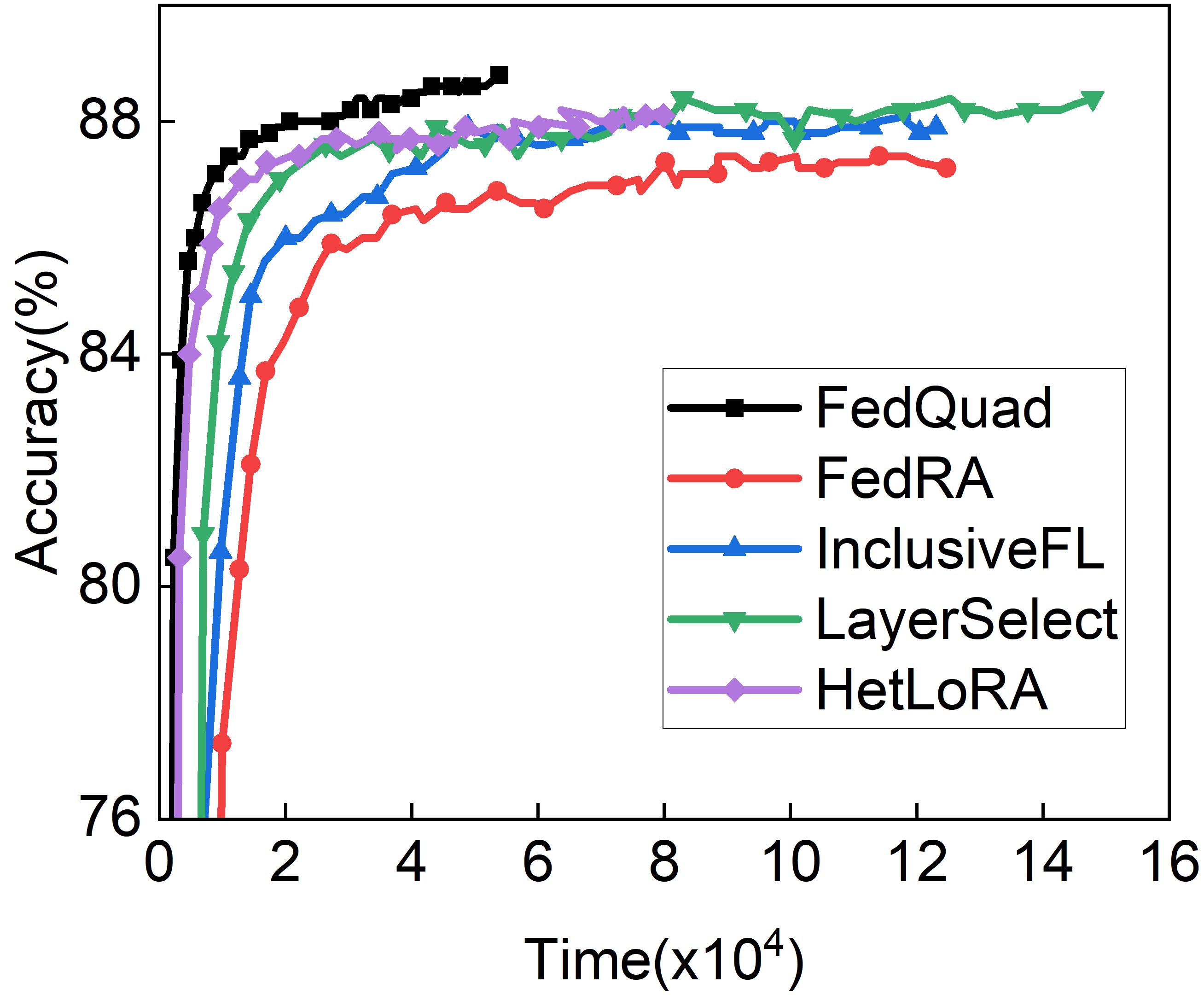} 
		\end{minipage}
		\label{time_with_acc_mnli}
	}
    	\subfigure[QQP]{
    		\begin{minipage}[b]{0.23\textwidth}
   		 	\includegraphics[width=1\textwidth]{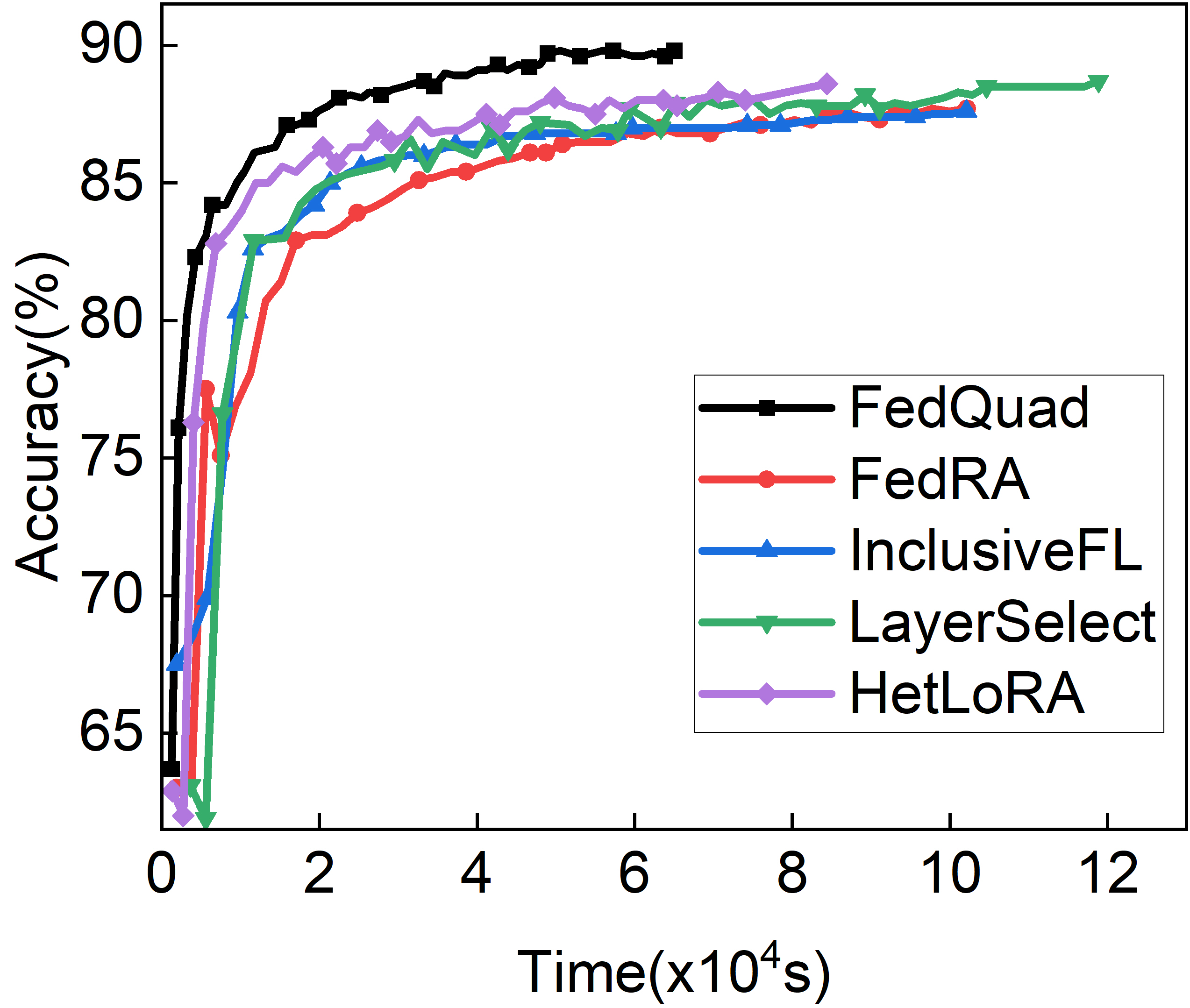}
    		\end{minipage}
		\label{time_with_acc_qqp}
    	}
	\subfigure[QNLI]{
		\begin{minipage}[b]{0.23\textwidth}
			\includegraphics[width=1\textwidth]{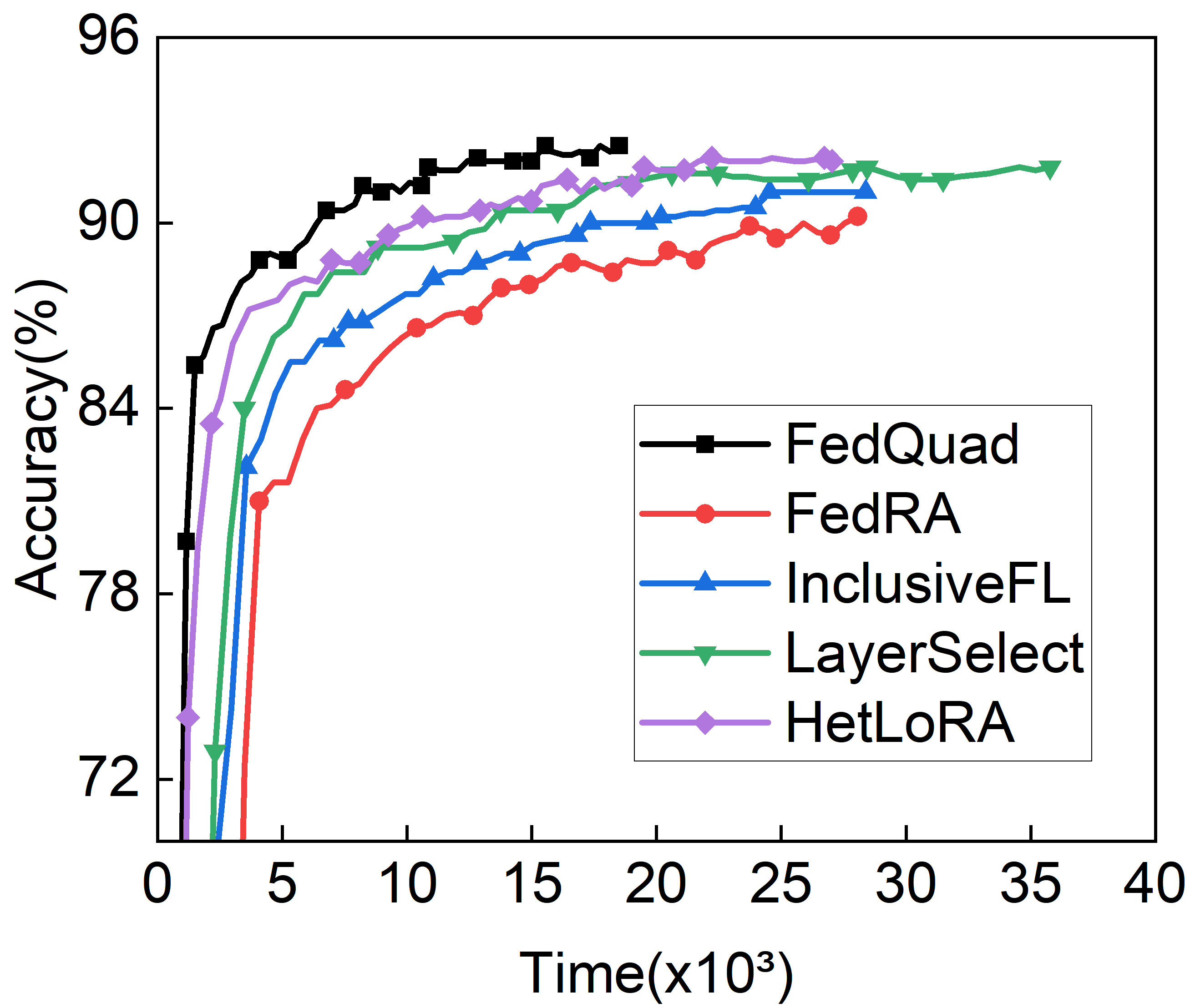} 
		\end{minipage}
		\label{time_with_acc_qnli}
	}
    	\subfigure[SST-2]{
    		\begin{minipage}[b]{0.23\textwidth}
		 	\includegraphics[width=1\textwidth]{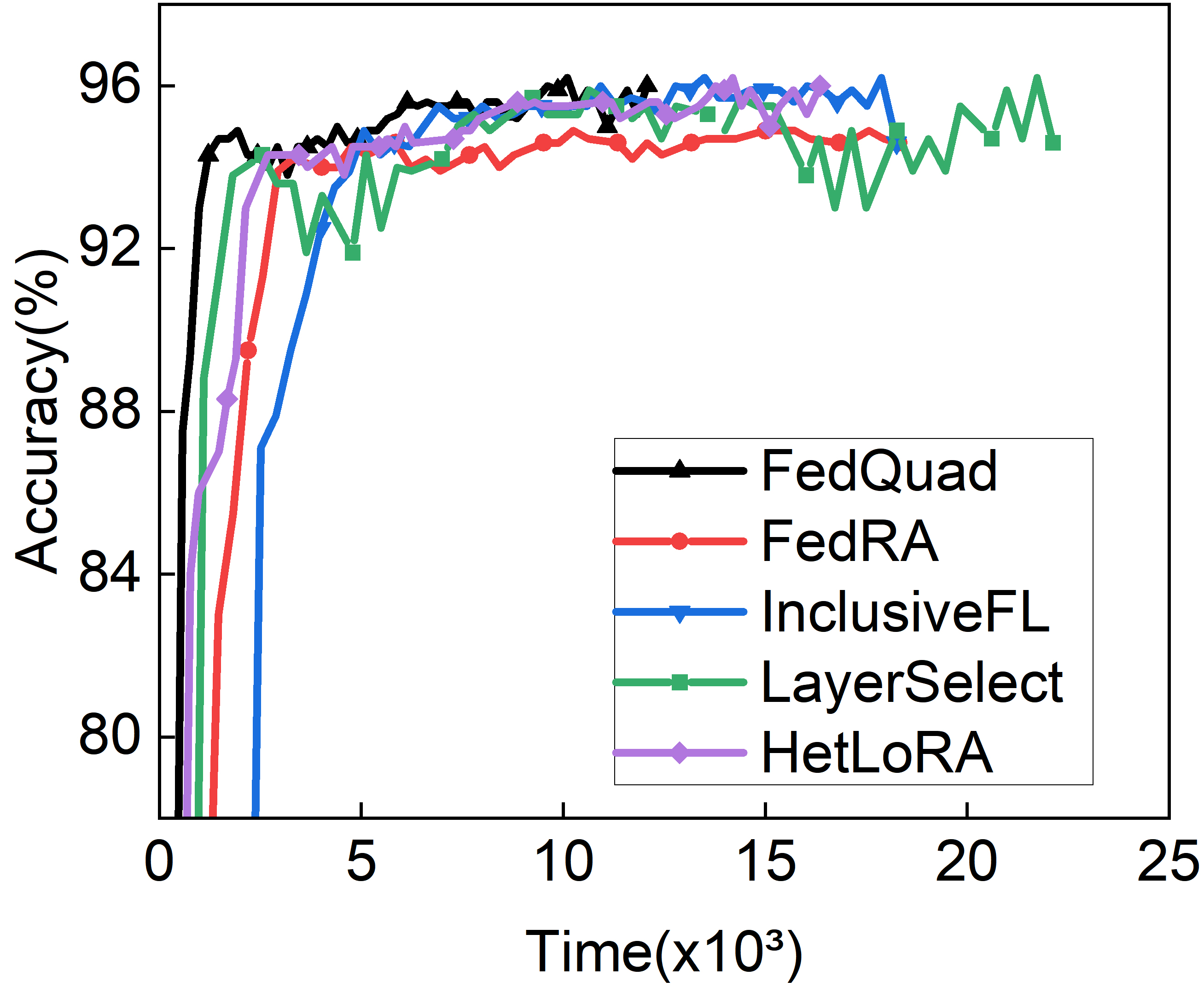}
    		\end{minipage}
		\label{time_with_acc_sst}
    	}
        \vspace{-0.4cm}
	\caption{Test accuracy of four approaches on the four datasets.}
        \vspace{-0.3cm}
        \label{time-acc}
\end{figure*}

\begin{figure*}
	\centering
	\subfigure[Time to reach 87\% accuracy]{
		\begin{minipage}[b]{0.23\textwidth}
			\includegraphics[width=1\textwidth]{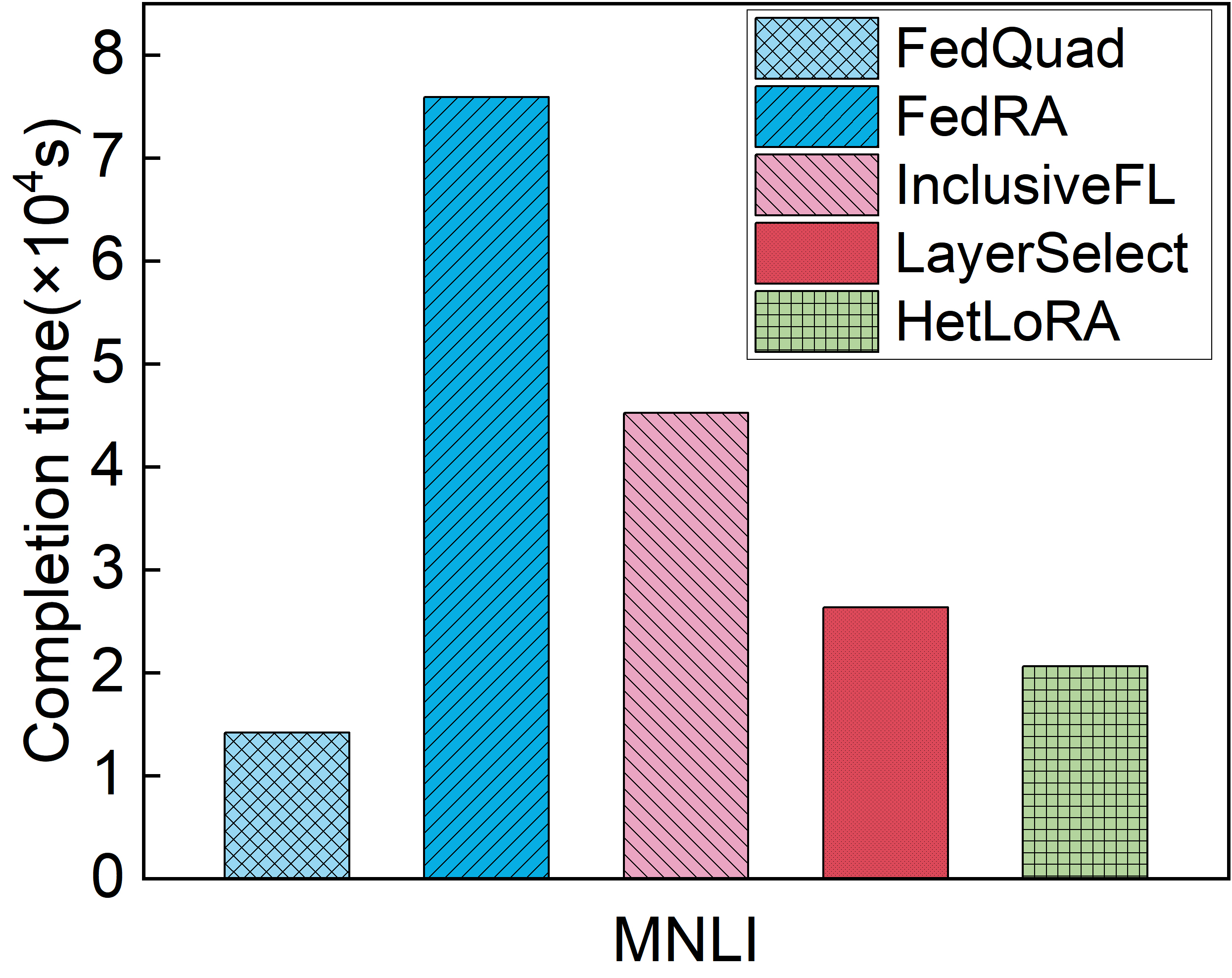} 
		\end{minipage}
		\label{roberta-completion-mnli}
	}
    	\subfigure[Time to reach 87\% accuracy]{
    		\begin{minipage}[b]{0.23\textwidth}
   		 	\includegraphics[width=1\textwidth]{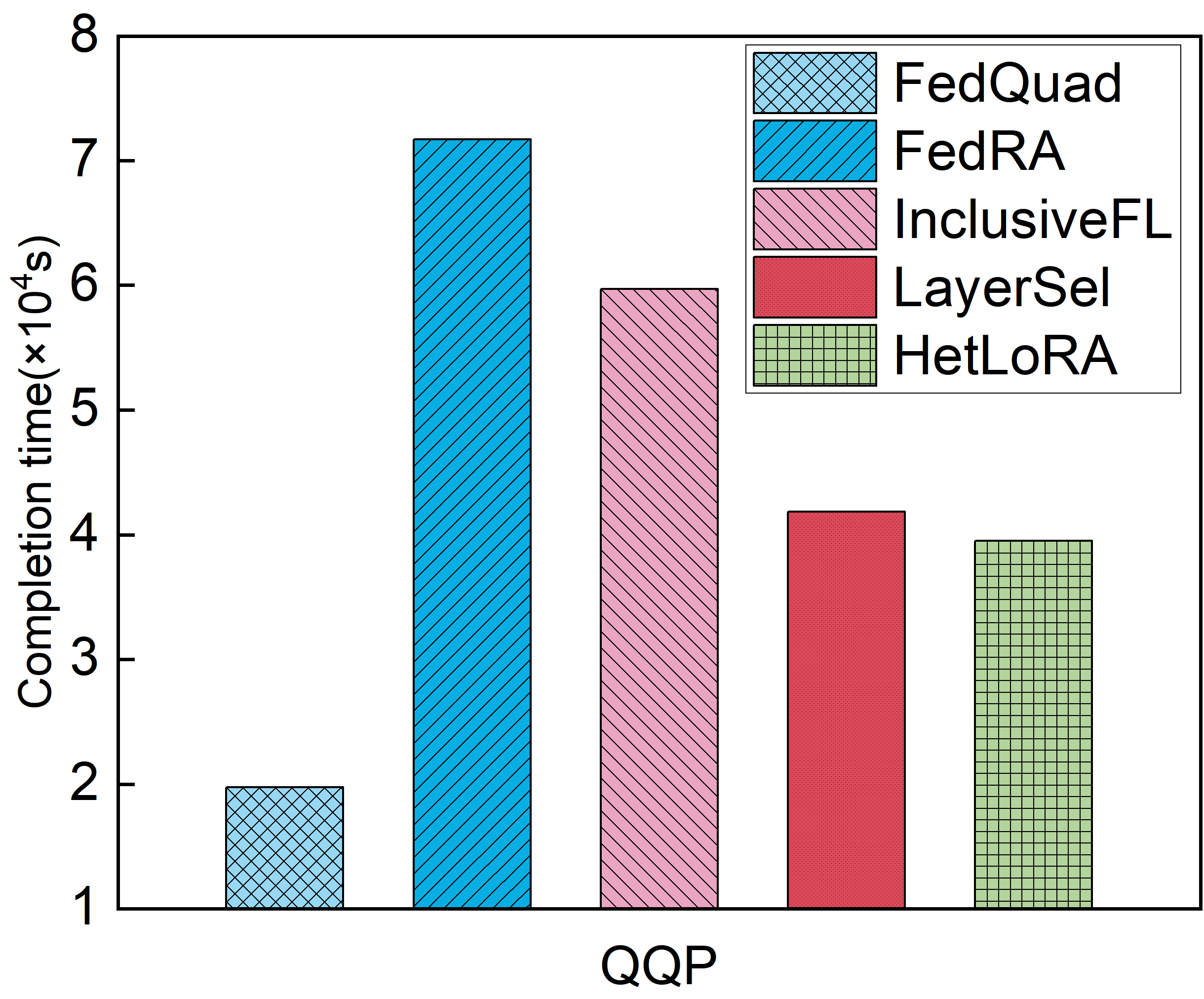}
    		\end{minipage}
		\label{roberta-completion-qqp}
    	}
	\subfigure[Time to reach 90\% accuracy]{
		\begin{minipage}[b]{0.23\textwidth}
			\includegraphics[width=1\textwidth]{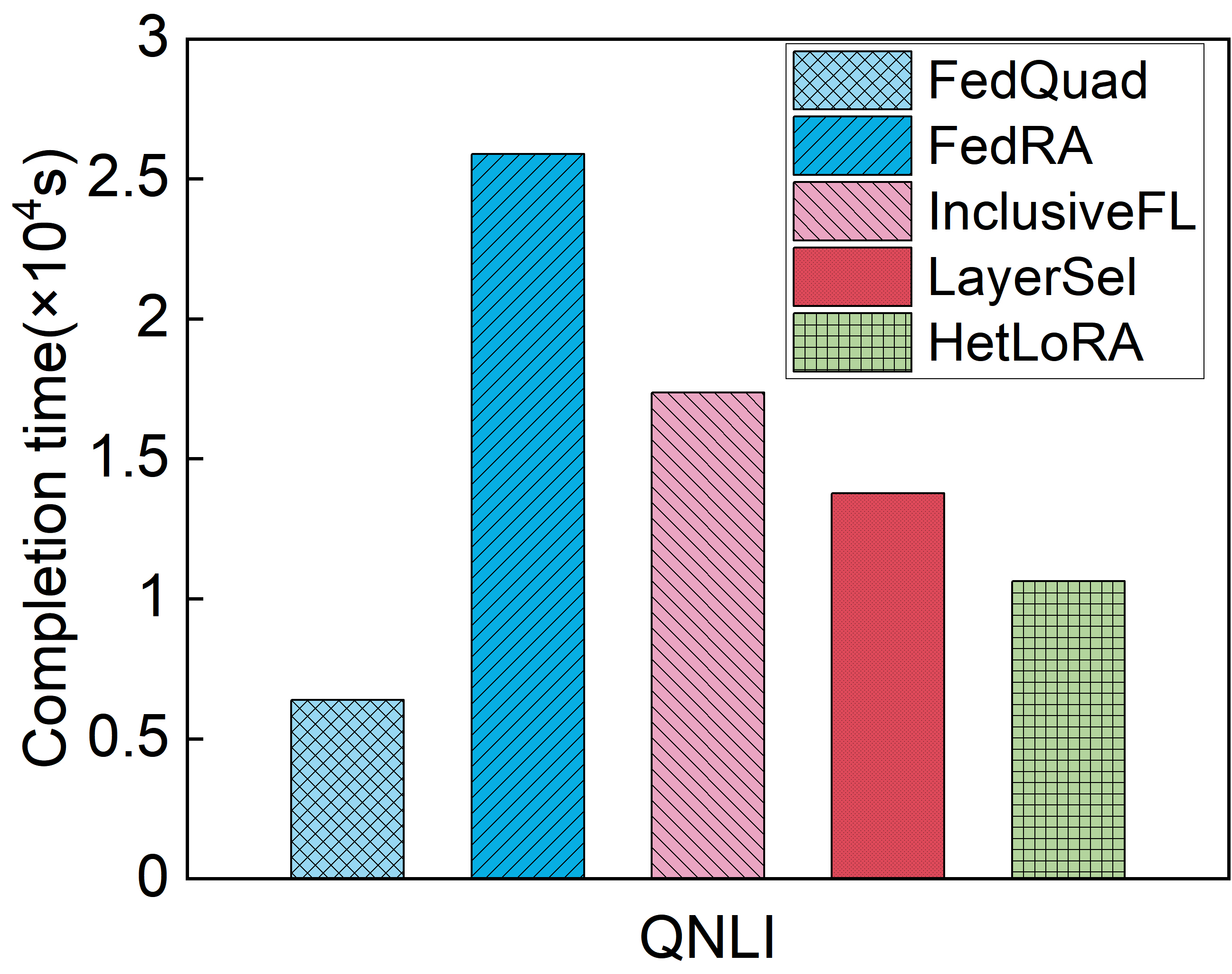} 
		\end{minipage}
		\label{roberta-completion-qnli}
	}
    	\subfigure[Time to reach 95\% accuracy)]{
    		\begin{minipage}[b]{0.23\textwidth}
		 	\includegraphics[width=1\textwidth]{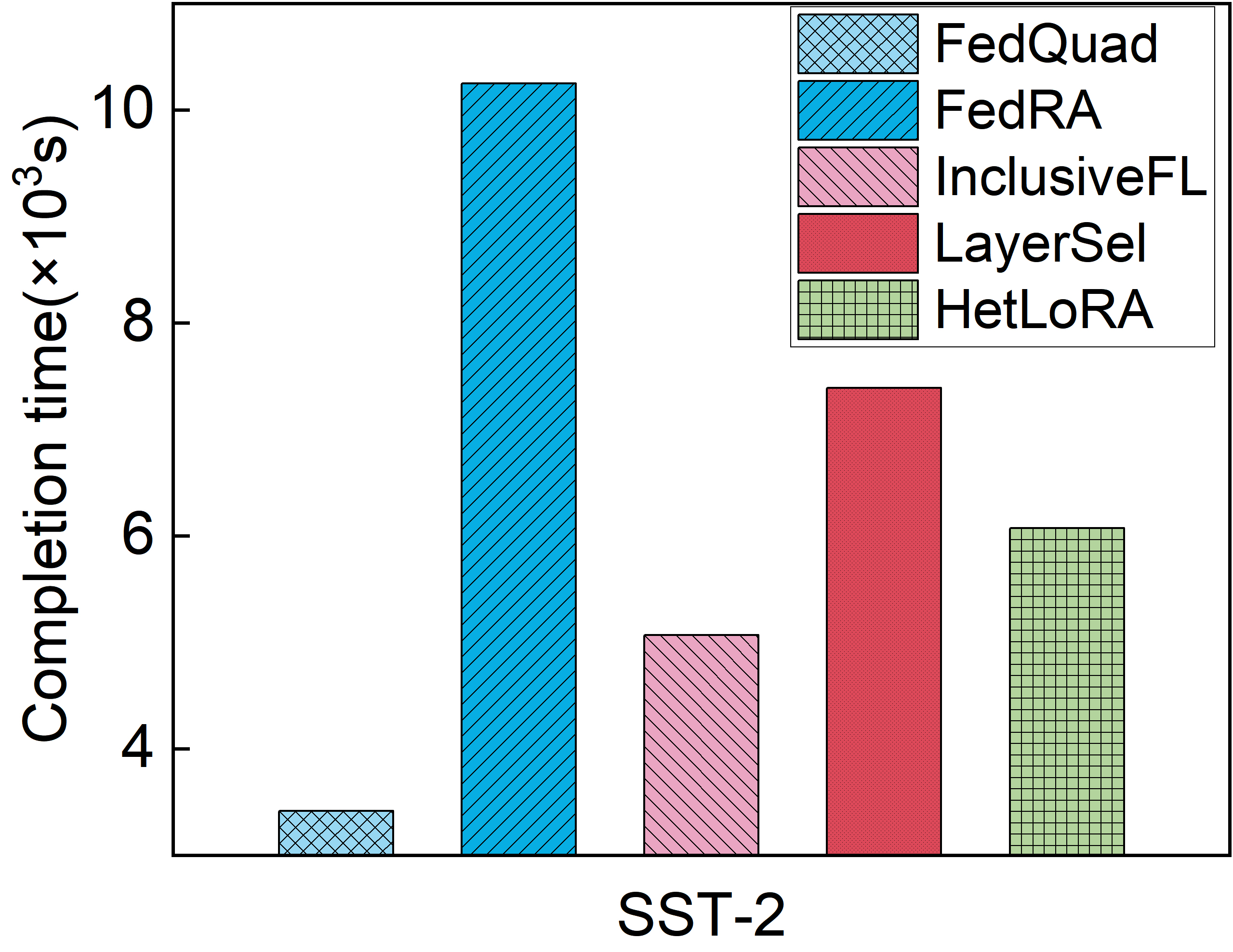}
    		\end{minipage}
		\label{roberta-completion-sst-2}
    	}
        \vspace{-0.4cm}
	\caption{Completion time of four approaches when achieving the target accuracy.}
        \vspace{-0.3cm}
        \label{completion-time}
\end{figure*}

\textbf{Baselines.} 
To evaluate the effectiveness of FedQuad, we compare it against four baseline methods: 
(1) FedRA \cite{su2024fedra} randomly selects subsets of transformer layers based on device resource constraints to construct sub-models and fine-tunes these sub-models using LoRA.  
(2) InclusiveFL \cite{liu2022no} allocates sub-models with varying numbers of consecutively stacked layers starting from the input layer, rather than selecting layers randomly, based on each device's resource constraints. To mitigate gradient loss in shallow models, it applies momentum distillation.
(3) LayerSel \cite{sun2024exploring} leverages local gradient information to select specific layers for fine-tuning across devices, while freezing the remaining layers instead of discarding them to conserve resources. Fine-tuning is then performed using LoRA.
(4) HetLoRA \cite{cho2024heterogeneous} allows devices to fine-tune heterogeneous local models by allocating different LoRA ranks to tackle system heterogeneity.

\textbf{Metrics.}
The following metrics are adopted to evaluate the performance of FedQuad and the baselines.

1) \textbf{Test Accuracy} reflects the accuracy of the models trained by different approaches on the test datasets, measured by the proportion of correctly predicted data.
Specifically, we record the test accuracy of the global model (the model after aggregation at the PS) in each round.

2) \textbf{Time-to-Accuracy} represents the total wall-clock time required for training a model to achieve a target accuracy . 
For fair comparisons, we set the target accuracy as the highest achievable accuracy by
FedQuad and all baselines. 
We record the completion time of each round, summing it up to obtain the total training time.

3) \textbf{Average Waiting Time} represents the average time each edge device spends waiting for global aggregation in each round, indicating the training efficiency of each method.

\textbf{Experimental Parameters.}
By default, all experiments are carried out on our prototype system and run 50 rounds.
Each device fine-tunes 1 epoch per round using AdamW~\cite{kingma2014adam} optimizer locally. The learning rate is set as 0.001 and decays according to a cosine scheduler. The batch size is fixed at 32 and the maximum sequence length is set to 128 for all experiments.


\subsection{Numerical Results}
 
\begin{figure*}
	\centering
	\subfigure[MNLI]{
		\begin{minipage}[b]{0.23\textwidth}
			\includegraphics[width=1\textwidth]{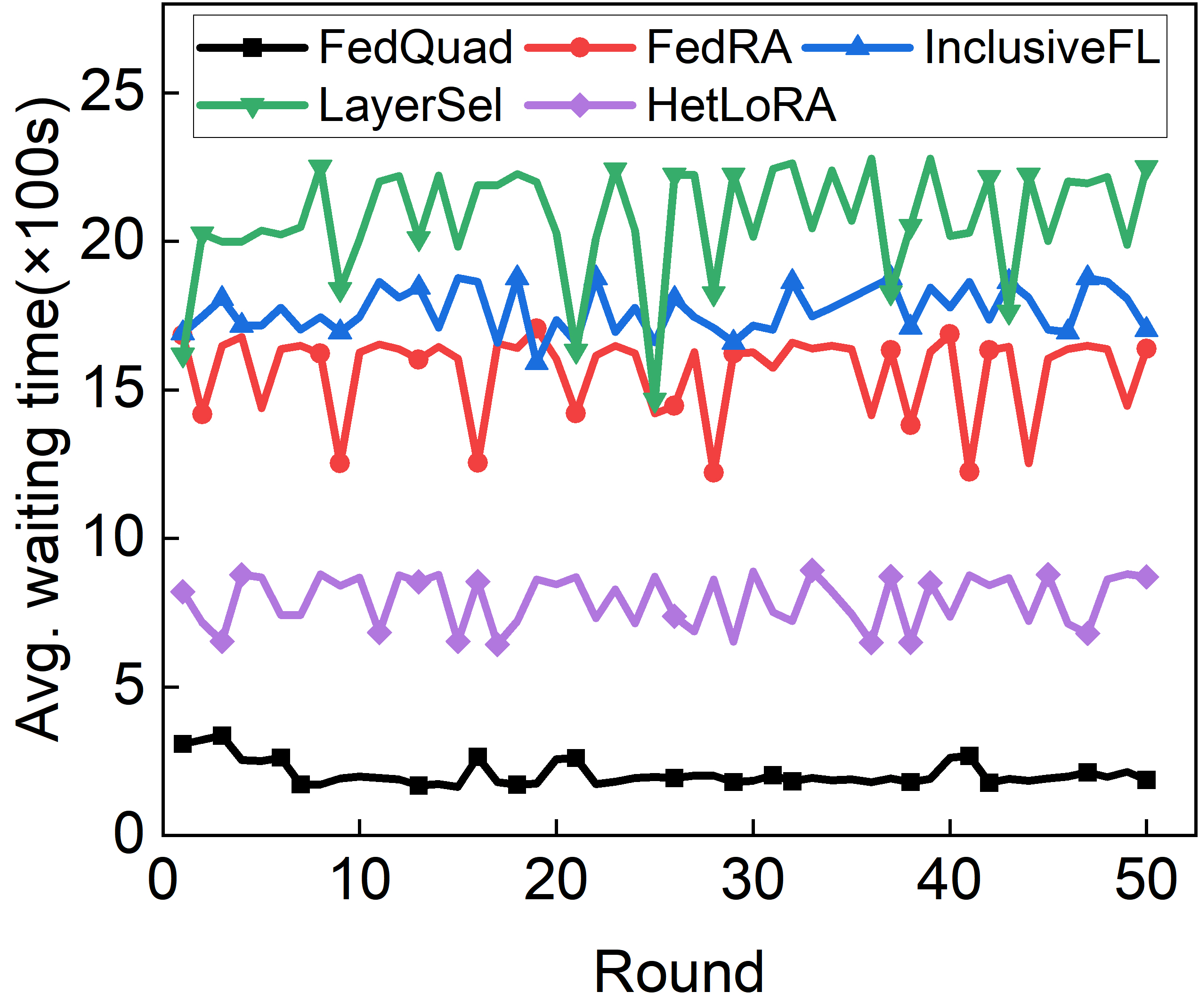} 
		\end{minipage}
		\label{roberta-waiting-mnli}
	}
    	\subfigure[QQP]{
    		\begin{minipage}[b]{0.23\textwidth}
   		 	\includegraphics[width=1\textwidth]{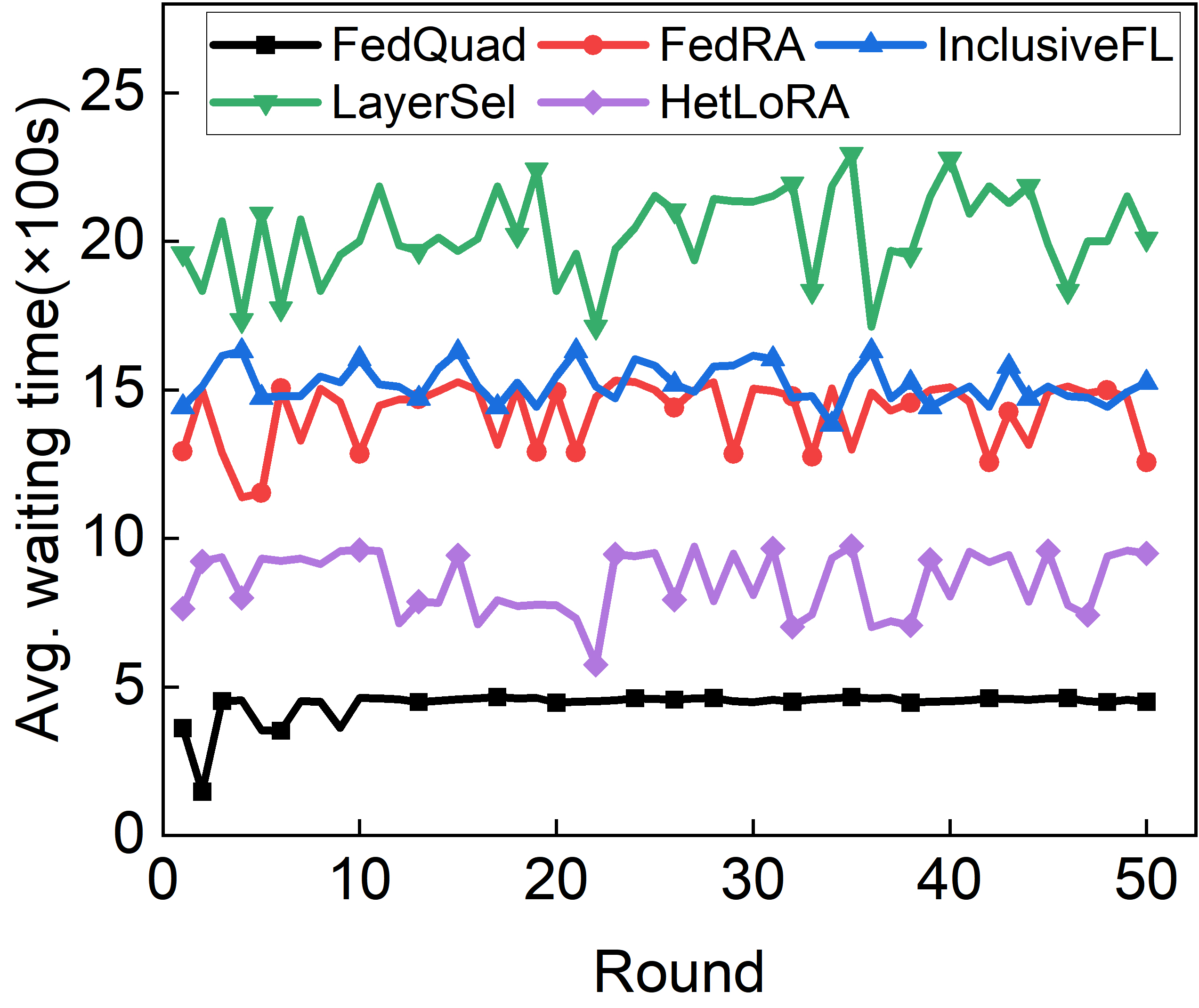}
    		\end{minipage}
		\label{roberta-waiting-qqp}
    	}
	\subfigure[QNLI]{
		\begin{minipage}[b]{0.23\textwidth}
			\includegraphics[width=1\textwidth]{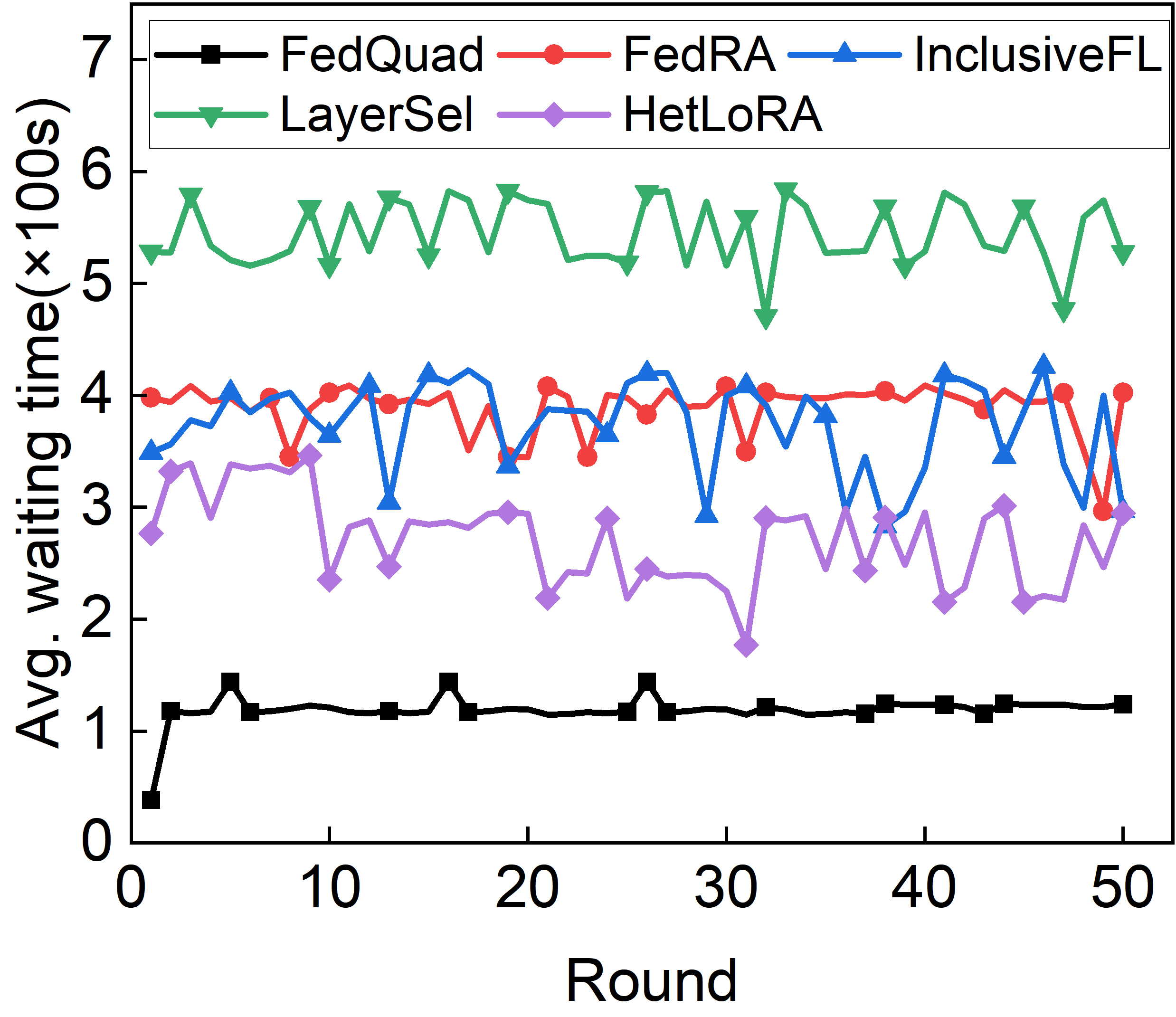} 
		\end{minipage}
		\label{roberta-waiting-qnli}
	}
    	\subfigure[SST-2]{
    		\begin{minipage}[b]{0.23\textwidth}
		 	\includegraphics[width=1\textwidth]{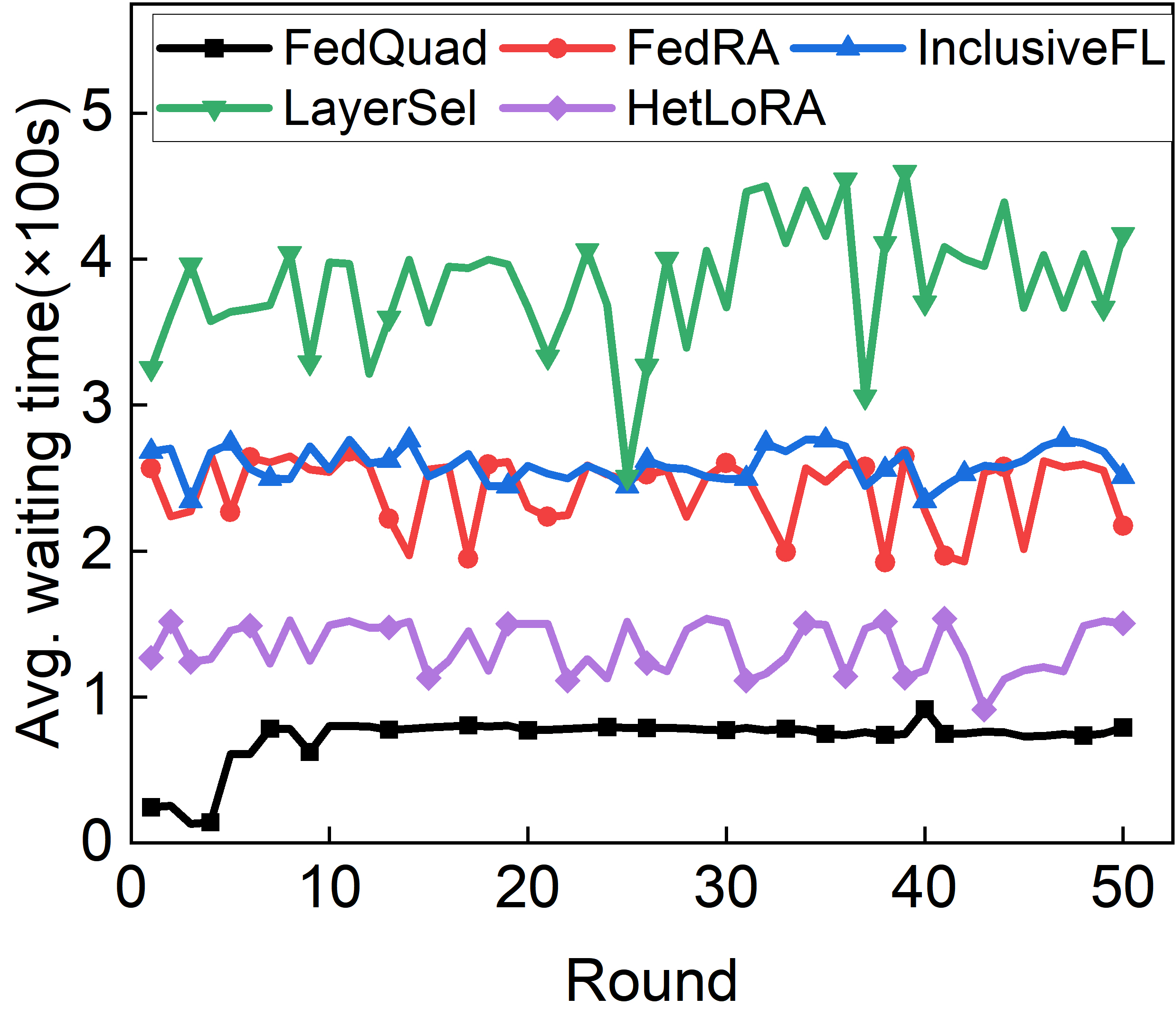}
    		\end{minipage}
		\label{roberta-waiting-sst-2}
    	}
        \vspace{-0.4cm}
	\caption{Average waiting time of four approaches on the four datasets.}
        \label{average-waiting-time}
        \vspace{-0.3cm}
\end{figure*}

\textbf{Overall Performance.} We conducte extensive experiments to evaluate the performance of FedQuad against several representative baselines. Figs.~\ref{time-acc} and~\ref{completion-time} illustrate the fine-tuning trajectories and overall completion time on various general language understanding tasks.

The results demonstrate that FedQuad achieves the fastest convergence speed across all tasks, significantly outperforming baselines. By adaptively allocating LoRA depth and quantized activation layers based on device-specific resource constrains, FedQuad enhances fine-tuning performance while reducing local training time. For instance, as shown in Figs.~\ref{time_with_acc_mnli} and~\ref{roberta-completion-mnli}, FedQuad reaches 87.5\% accuracy on MNLI within only 14,213 s, whereas FedRA, InclusiveFL, LayerSel, and HetLoRA require 75,928s, 45,262s, 26,350s, and 20,642s, respectively. Compared to FedRA, InclusiveFL, LayerSel, and HetLoRA, FedQuad provides 5.3$\times$, 3.2$\times$, 1.85$\times$, and 1.45$\times$ speedup.
Similarly, Figs.~\ref{time_with_acc_qqp} and~\ref{roberta-completion-qqp} show that FedQuad achieves 87\% accuracy on QQP in 19,738s, while FedRA, InclusiveFL, LayerSel, and HetLoRA take 71,707s, 59,700s, 41,845s, and 32575s, respectively. The corresponding speedups are 3.6$\times$, 3.0$\times$, 2.1$\times$, and 1.65$\times$. Moreover, Figs.~\ref{time-acc} and~\ref{completion-time} further indicate that FedQuad consistently achieves superior completion times on QNLI and SST-2 as well. 
As shown in Table \ref{table-time-comparison}, FedQuad consistently achieves the fastest convergence across all evaluated models, significantly outperforming baseline methods. For instance, on BERT-large, FedQuad reduces convergence time by 80.2\%, 83.0\%, 53.5\%, and 28.8\% compared to FedRA, InclusiveFL, LayerSel, and HetLoRA, respectively. Similarly, on DeBERTa-large, FedQuad attains convergence time reductions of 79.5\%, 73.9\%, 34.9\%, and 22.4\%, respectively. 
These results collectively confirm FedQuad’s advantage in accelerating the fine-tuning process.

These results demonstrate that FedQuad achieves both improved fine-tuning performance and reduced convergence time. Unlike FedRA and InclusiveFL, which reduce resource consumption by directly skipping transformer layers, FedQuad employs a layer freezing strategy. This strategy avoids undesirable side effects such as output bias and accuracy degradation that often arise from layer dropping during forward propagation. Moreover, in both FedRA and InclusiveFL, memory limitations may hinder stronger devices from fully exploiting their computational capabilities, leaving them idle as they wait for slower devices, thereby reducing overall fine-tuning efficiency and prolonging convergence time.
Compared to more competitive baselines like LayerSel and HetLoRA, FedQuad also exhibits notable advantages. LayerSel selects layers for fine-tuning based on gradient norms, but the uneven distribution of resource consumption during fine-tuning leads to slower convergence and presents deployment challenges, as memory-constrained devices often cannot afford to compute gradient norms for all layers. HetLoRA, on the other hand, adapts LoRA ranks to device capabilities, but fails to fundamentally address the inherent resource overhead of PEFT. Adjusting only the LoRA rank is insufficient to overcome the system bottlenecks.

In contrast, FedQuad comprehensively models real-world device resource constrains. It employs a back-to-front layer selection strategy for fine-tuning, combined with activation quantization to mitigate memory constraints and fully leverage computational resources. This enables adaptive configuration of both LoRA depth and quantization layers for each device based on its resource constrains. As a result, FedQuad achieves substantially improved fine-tuning performance and significantly faster convergence speed.

\begin{table}[htbp]
\centering
\footnotesize
\renewcommand{\arraystretch}{1.3}
\setlength{\tabcolsep}{2pt} 
\resizebox{0.48\textwidth}{!}{ 
\begin{tabular}{ccccccc}
\toprule
\textbf{Model} & \textbf{Target Acc} 
& \textbf{FedQuad} & \textbf{FedRA} 
& \textbf{InclusiveFL} & \textbf{LayerSel} & \textbf{HetLoRA} \\
\midrule
BERT-large     & 81\% & 14.1 & 71.1 & 83.1 & 30.3  & 19.8 \\
DeBERTa-large  & 88\% & 23.5 & 114.4 & 90.0 & 36.1 & 30.3 \\
RoBERTa-large  & 87\% & 14.2  & 75.9 & 45.2 & 26.3 & 20.6 \\
\bottomrule
\end{tabular}
}
\caption{Comparison of  completion time (in $10^3$ s) across models on the MNLI dataset using different baselines.}
\label{table-time-comparison}
\end{table}

\begin{table}[t]
  \centering\footnotesize
  \setlength{\tabcolsep}{6pt}
  \renewcommand{\arraystretch}{1.2}

  \resizebox{\columnwidth}{!}{%
  \begin{tabular}{l|cc|cc|cc|}
    \toprule
    \multicolumn{1}{c|}{\textbf{Heter. Level}} &
      \multicolumn{2}{c|}{\textbf{Low}} &
      \multicolumn{2}{c|}{\textbf{Medium}} &
      \multicolumn{2}{c|}{\textbf{High}} \\
    \midrule
    \textbf{Target Acc} &
      \multicolumn{2}{c|}{88.1} &
      \multicolumn{2}{c|}{88.0} &
      \multicolumn{2}{c|}{87.5} \\
    \midrule
    \textbf{Metric} & \textbf{Time} & \textbf{Final Acc} &
                     \textbf{Time} & \textbf{Final Acc} &
                     \textbf{Time} & \textbf{Final Acc} \\
    \midrule
    FedQuad         &  5.2& 88.9&  8.7
& 88.7& 14.2 & 88.8\\
    FedRA           &  7.8& 88.1& 21.4
& 88.0
& 75.9 & 87.5\\
    InclusiveFL     &  7.2& 88.3&  15.7
& 88.2
& 45.2 & 87.9\\
    LayerSel        &  6.5& 88.6&  14.6& 88.4
& 26.3 & 88.4\\
    HetLoRA         &  6.1& 88.5&  11.2& 88.5& 20.6 & 88.1\\
    \bottomrule
  \end{tabular}}
  \caption{Comparison of completion time (in $10^{3}$ s) and final accuracy (\%) on MNLI under different heterogeneity levels.}
  \label{table-time-acc-final}
\end{table}

\textbf{Effect of Device Heterogeneity.}
To comprehensively assess FedQuad’s effectiveness under varying degrees of device heterogeneity, we evaluate FedQuad alongside baseline methods across three heterogeneity levels, \ie, Low, Medium, and High. Specifically, under the Low heterogeneity setting, we configure all devices as high-performance devices. For Medium, we equally distribute devices between high-performance and moderate-performance categories at a ratio of 1:1. In the High heterogeneity scenario, we assign the device proportions as 3:3:4 for strong-, moderate-, and weak-performance devices, respectively. As shown in Table \ref{table-time-acc-final}, we observe that as device heterogeneity increases, the time required by all methods to reach the target accuracy grows significantly, which aligns with expectations when more resource-constrained devices are introduced into the system. However, compared with the baseline methods, FedQuad demonstrates remarkable advantages across all heterogeneity levels. For example, on the MNLI dataset under medium heterogeneity, FedQuad achieves speedups of approximately
$2.46\times$, $1.80\times$, $1.68\times$, and $1.29\times$ over FedRA, InclusiveFL, LayerSel, and HetLoRA, respectively. Moreover, as heterogeneity rises from low to high, the performance gap between FedQuad and the other baselines further widens: FedQuad attains a $1.5\times$ speedup over FedRA in low heterogeneity scenarios, which increases to $5.35\times$ in high heterogeneity scenarios.  These results indicate that FedQuad maintains robust and efficient fine-tuning capabilities across varying degrees of device heterogeneity.

To further assess FedQuad’s capability in handling system heterogeneity, we report the average waiting time across four datasets in Fig.~\ref{average-waiting-time}.  FedQuad achieves the lowest average waiting time across all datasets. For example, on MNLI as illustrated in Fig.~\ref{roberta-waiting-mnli}, FedQuad achieves the shortest average waiting time in each round. Furthermore, over all rounds, FedQuad reduces the average waiting time by 68.9\%, 70.9\%, 78.2\%, and 48.1\% compared to FedRA, InclusiveFL, LayerSel, and HetLoRA, respectively. On QQP, FedQuad achieves an average waiting time of 442.5s, while the other baselines require 1423.1s, 1520.4s, 2026.1s, and 852.4s, corresponding to reductions of 68.9\%, 70.9\%, 78.2\%, and 51.9\%, respectively. This improvement stems from the fact that, although LayerSel selects the most important layers for fine-tuning, those layers can incur high computational overhead and thus long synchronization delays. FedRA and InclusiveFL adapt to heterogeneous devices by discarding layers but fail to fully leverage the compute capacity of more powerful devices under tight memory constraints, resulting in prolonged waiting times. HetLoRA mitigates heterogeneity by assigning devices to different LoRA levels but achieves only limited gains. In contrast, FedQuad adaptively determines the optimal combination of LoRA depth and activation quantization layers for each device based on its compute and memory resources, thereby maximizing resource utilization and minimizing synchronization delays for efficient fine-tuning.

\vspace{-0.2cm}
\subsection{Ablation Study}
\label{sec-ablation}

FedQuad incorporates two critical factors, \ie, LoRA depth and the number of activation quantization layers, which are specifically designed to enhance the efficiency and adaptability of FedLoRA in heterogeneous environments. In this section, we perform a series of ablation experiments on the MNLI and QQP datasets to evaluate the individual contributions of these two factors.

\begin{figure}[t]
    \centering
    \subfigure[MNLI]{
        \begin{minipage}[t]{0.5\linewidth}
        \centering
        \includegraphics[width=1.5in]{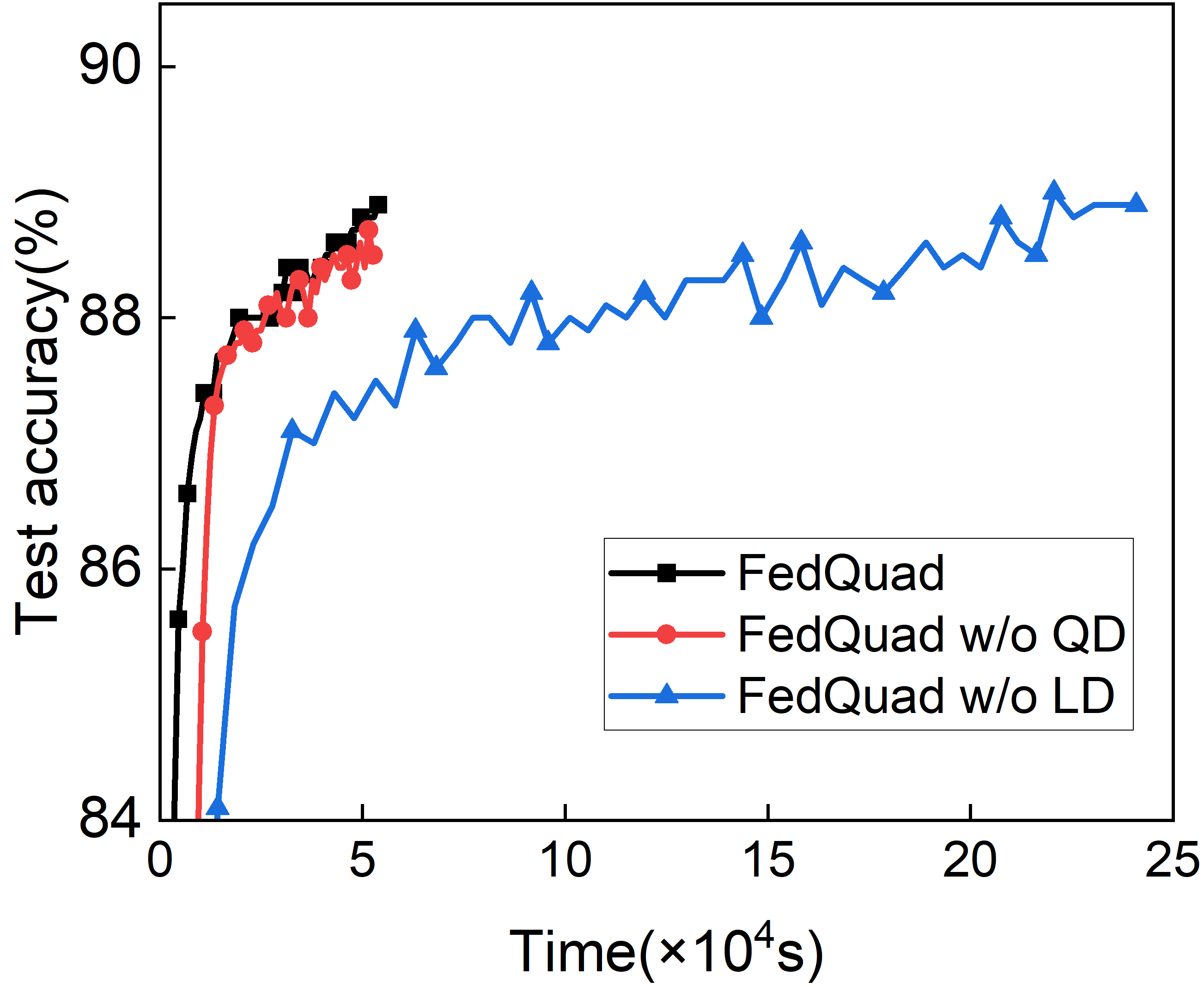}
        \end{minipage}%
        \label{ablation-mnli}
    }%
    \subfigure[QQP]{
        \begin{minipage}[t]{0.5\linewidth}
        \centering
        \includegraphics[width=1.5in]{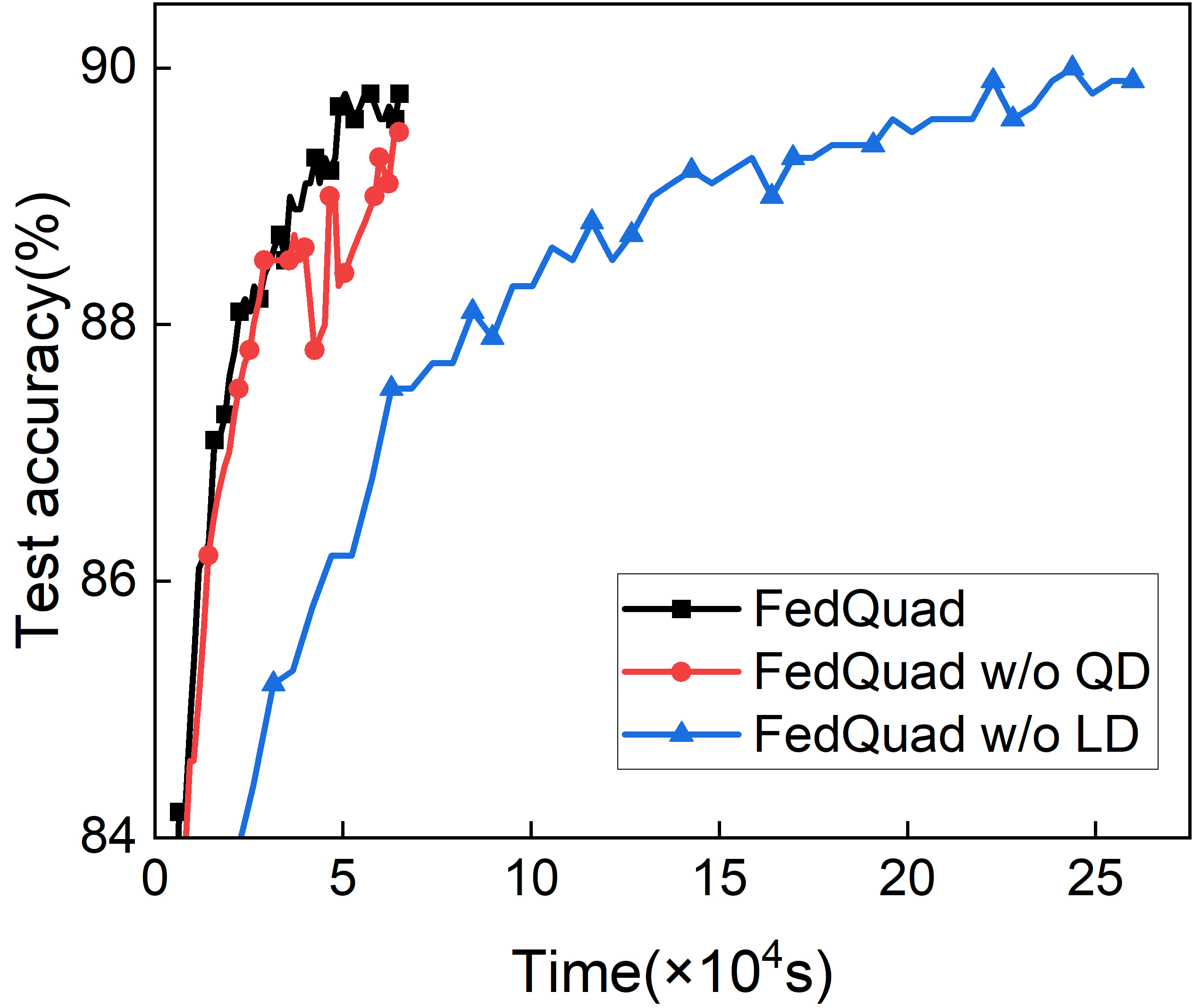}
        \end{minipage}%
        \label{ablation-qqp}
    }%
    \centering
    \vspace{-0.25cm}
    \caption{Effect of LoRA depth and activation quantization.}
    \label{ablation-fig}
    \vspace{-0.5cm}
\end{figure}

We compare three variants: (i) the full version of our method (FedQuad), (ii) a version without activation quantization (FedQuad w/o QD), and (iii) a version that applies the maximum possible number of activation quantization layers under memory constraints but removes adaptive LoRA depth (FedQuad w/o LD). As shown in Fig.~\ref{ablation-fig}, both LoRA depth and activation quantization play essential roles in FedQuad, though they affect system performance in different ways.

For instance, as illustrated in Fig.~\ref{ablation-fig}, both the FedQuad w/o QD and FedQuad w/o LD variants achieve final test accuracies comparable to that of FedQuad. However, when targeting an $88\%$ accuracy on MNLI, FedQuad reduces training completion time by approximately $24.5\%$ compared to FedQuad w/o QD and by approximately $74.7\%$ compared to FedQuad w/o LD. Similarly, on QQP, FedQuad accelerates the fine-tuning process by factors of $1.29\times$ and $3.53\times$ relative to FedQuad w/o QD and FedQuad w/o LD, respectively, when reaching the same $89\%$ accuracy target.

These observations highlight the distinct roles and complementary benefits of the two factors. On the one hand, FedQuad w/o LD, which retains activation quantization, allows devices to fine-tune a larger number of LoRA layers, thereby improving final accuracy but at the cost of increased waiting time and slower convergence. On the other hand, FedQuad w/o QD, which removes activation quantization, restricts faster devices from utilizing their full capacity to fine-tune additional layers, limiting potential performance gains.

Overall, these results clearly validate the necessity of integrating both adaptive LoRA depth and activation quantization into FedQuad to achieve efficient and effective fine-tuning under heterogeneous system constraints. 

\section{Related Work}\label{sec:related}

\textbf{Federated Fine-Tuning of LLMs.}
To fully leverage data on end devices while preserving user privacy, federated fine-tuning (FedFT) has emerged as a promising paradigm for adapting large language models (LLMs) to decentralized settings. For example, FedNLP \cite{lin2021fednlp} provides a benchmark framework for evaluating FedFT across various NLP tasks. To address communication overhead on devices, recent studies have turned to parameter-efficient fine-tuning (PEFT) techniques. Notably, FedAdapter~\cite{zhang2023fedpetuning} employs lightweight adapter modules to accelerate model convergence in FL, while FedLoRA integrates low-rank adapters~\cite{hu2021lora} to improve training speed and accuracy.

\textbf{Resource-efficient Approaches.}
Although PEFT methods help reduce communication overhead, fine-tuning LLMs on end devices still suffers from severe memory and computational limitations. To adapt to these constraints, existing works explore various fine-tuning strategies to tailor model architectures or training processes to device-specific resource capabilities. For instance, FedRA~\cite{su2024fedra} constructs sub-models by randomly selecting subsets of transformer layers based on device resource availability and fine-tunes them with LoRA. However, directly discarding layers compromises the model architecture and degrades fine-tuning accuracy. Alternatively, LayerSel~\cite{sun2024exploring} selects specific layers to fine-tune using local gradient information, freezing the rest rather than removing them to conserve memory. Yet, it overlooks the fact that resource consumption is unevenly distributed across layers, which limits its practical deployability. HetLoRA~\cite{cho2024heterogeneous} tackles system heterogeneity by assigning different LoRA ranks to each device, enabling them to fine-tune heterogeneous local models. However, merely adjusting the LoRA rank is insufficient to fundamentally address the resource constraints in FedFT.

\section{Conclusion}\label{sec:conclusion}
In this paper, we have proposed FedQuad, a novel and efficient federated fine-tuning framework designed to overcome the challenges posed by resource constraints and system heterogeneity. Specifically, FedQuad adaptively adjusts the LoRA depth to match the capabilities of individual devices and employs activation quantization to alleviate memory overhead, thus facilitating efficient deployment of large language models on resource-limited devices. Additionally, we have analyzed the interplay between LoRA depth and the number of activation quantization layers, and devised a greedy-based algorithm to jointly determine optimal configurations for heterogeneous devices, significantly enhancing fine-tuning efficiency. Extensive experimental results demonstrate that FedQuad achieves superior convergence performance and accelerates fine-tuning by $1.4$–$5.3\times$ compared to state-of-the-art baselines.


\bibliographystyle{IEEEtran}
\bibliography{content/refs}

\end{document}